\documentclass[twocolumn]{aastex631}
\shorttitle{Gaussian Processes for QPO Detection}
\shortauthors{H\"ubner et al.}

\usepackage{amssymb}
\usepackage{amsmath}
\usepackage{graphicx}
\usepackage{dcolumn}
\usepackage{xspace}

\usepackage{subfigure}
\let\tablenum\relax
\usepackage{siunitx}
\DeclareSIUnit \parsec {pc}

\usepackage[normalem]{ulem} 

\newcommand{\bilby}{\textsc{Bilby}\xspace}

\newcommand{\celerite}{\textsc{celerite}\xspace}

\newcommand{\dynesty}{\textsc{dynesty}\xspace}

\newcommand{\Z}{\mathcal{Z}}
\newcommand{\M}{\mathcal{M}}
\renewcommand{\L}{\mathcal{L}}
\newcommand{\BF}{\text{BF}}

\newcommand{\arn}{a_{\mathrm{rn}}}
\newcommand{\aqpo}{a_{\mathrm{qpo}}}
\newcommand{\crn}{c_{\mathrm{rn}}}
\newcommand{\cqpo}{c_{\mathrm{qpo}}}
\newcommand{\fqpo}{f_{\mathrm{qpo}}}
\newcommand{\krn}{k_{\mathrm{rn}}}
\newcommand{\kqpo}{k_{\mathrm{qpo}}}
\newcommand{\krnqpo}{k_{\mathrm{qpo+rn}}}
\newcommand{\BFQPO}{BF_{\mathrm{qpo}}}

\newcommand{\MelbUni}{School of Physics, University of Melbourne, Parkville, VIC 3010, Australia}
\newcommand{\OzGrav}{Australian Research Council Centre of Excellence for Gravitational Wave Discovery (OzGrav)}
\newcommand{\SPA}{School of Physics and Astronomy, Monash University, Clayton, VIC 3800, Australia}
\newcommand{\SRON}{SRON Netherlands Institute for Space Research, Niels Bohrlaan 4, 2333 CA Leiden, The Netherlands}

\newcommand{\CCPP}{Center for Cosmology and Particle Physics, Department of Physics, New York University, 726 Broadway, New York, NY 10003}
\newcommand{\MARL}{Music and Audio Research Laboratory, New York University, 370 Jay St., Brooklyn, NY 11201}
\newcommand{\CUA}{Physics Department, The Catholic University of America, Washington, DC, 20064, USA}
\newcommand{\SPL}{Solar Physics Laboratory, NASA Goddard Space Flight Center, Greenbelt, MD, 20771, USA}

\begin{document}

\title{Searching for quasi-periodic oscillations in astrophysical transients using Gaussian processes}

\author{Moritz H\"ubner}
    \email{moritz.thomas.huebner@gmail.com}
    \affiliation{\MelbUni}
    \affiliation{\OzGrav}

\author{Daniela Huppenkothen}
    \email{d.huppenkothen@sron.nl}
    \affiliation{\SRON}

\author{Paul D. Lasky}
    \affiliation{\SPA}
    \affiliation{\OzGrav}

\author{Andrew R. Inglis}
    \affiliation{\CUA}
    \affiliation{\SPL}

\author{Christopher Ick}
    \affiliation{\MARL}

\author{David W. Hogg}
    \affiliation{\CCPP}

\begin{abstract}

Analyses of quasi-periodic oscillations (QPOs) are important to understanding the dynamic behaviour in many astrophysical objects during transient events like gamma-ray bursts, solar flares, magnetar flares and fast radio bursts.
Astrophysicists often search for QPOs with frequency-domain methods such as (Lomb-Scargle) periodograms, which generally assume power-law models plus some excess around the QPO frequency.
Time-series data can alternatively be investigated directly in the time domain using Gaussian Process (GP) regression.
While GP regression is computationally expensive in the general case, the properties of astrophysical data and models allow fast likelihood strategies.
Heteroscedasticity and non-stationarity in data have been shown to cause bias in periodogram-based analyses.
Gaussian processes can take account of these properties.
Using GPs, we model QPOs as a stochastic process on top of a deterministic flare shape.
Using Bayesian inference, we demonstrate how to infer GP hyperparameters and assign them physical meaning, such as the QPO frequency.
We also perform model selection between QPOs and alternative models such as red noise and show that this can be used to reliably find QPOs.
This method is easily applicable to a variety of different astrophysical data sets.
We demonstrate the use of this method on a range of short transients: a gamma-ray burst, a magnetar flare, a magnetar giant flare, and simulated solar flare data. 

\end{abstract}


\section{Introduction}

In the analysis of astrophysical time series, quasi-periodic oscillations (QPOs) are peaks of excess energy in the power spectrum of the time series, corresponding to semi-coherent oscillations where the amplitude, the oscillation frequencies or both may jitter.
Quasi-periodic oscillations have been observed in many different astrophysical sources such as giant magnetar flares~\citep{Israel2005, Strohmayer2005, Strohmayer2006, Watts2006, Huppenkothen2014a, Miller2019, Castro-Tirado2021}, X-ray binaries (see \citet{Ingram2019} for a review), and solar flares (See \citet{Nakariakov2009, Zimovets2021} for reviews).

They provide an exciting avenue to probe the properties of astrophysical objects.
For example, QPOs in giant magnetar flares are likely associated with magnetic coupling of the crust to the core~\citep{Levin2006, Levin2007, Colaiuda2009, Levin2011, Gabler2011, Gabler2013}.
Identifying torsional modes may eventually shed light on neutron star properties~\citep{Samuelsson2007, Sotani2007, Colaiuda2011, Sotani2016, Sotani2017}.
Quasi-periodic oscillations may also eventually uncover the nature and properties of the central engine of GRBs~\citep{Ziaeepour2011}.
In solar flares, it is not yet firmly established if QPOs originate from magnetohydrodynamic oscillations in coronal structures or due to magnetic reconnection~\citep{Zimovets2021}, and understanding these processes will also shed light on the origin of stellar flares.

Most common analyses of time series that look for QPOs use frequency-domain based methods such as periodograms or power spectra~\citep{Israel2005, Strohmayer2005, Strohmayer2006, Watts2006, DeLuca2010, Cenko2010, Huppenkothen2012, Huppenkothen2014a, Huppenkothen2014b, Inglis2015, Huppenkothen2015, Auchere2016, Inglis2016, Huppenkothen2017, Broomhall2019, Hayes2019, Miller2019, Hayes2020, Andersen2021, Castro-Tirado2021, Tarnopolski2021, PastorMarazuela2022}.
One performs analysis under the assumption that bins in a periodogram are $\chi^2_2$-distributed, and hence a Whittle likelihood applies.
However, this is strictly only true for infinitely long time series with homoscedastic stationary Gaussian data.
The assumption of stationarity in particular is an issue for rapidly varying, non-stationary time series such as those observed from fast transients like Gamma-ray Bursts (GRBs), magnetar bursts, and recently, Fast Radio Bursts (FRBs), where the total duration of the burst is perhaps only a few times longer than a candidate QPO period.
Previous work has shown that at for fast transients, the statistical distributions at low frequencies deviate significantly from those assumed for a stationary process \citep{Huppenkothen2012}, and that assuming a Whittle likelihood can in some circumstances severely overstate the actual significance of QPOs~\citep{Huebner2021}.
Specifically, if the QPO is not present for the entire time series, frequency bins around the central QPO frequency are not statistically independent, which leads to an overestimate of the QPO significance.
Furthermore, \citet{Auchere2016} has shown that detrending methods meant to eliminate the deterministic part of the time series cause artificial oscillatory behaviour, which might be confused for real QPOs.

An alternative to frequency-domain methods is to analyse the data in the time domain, which has some advantages over periodogram-based methods. 
When calculating a periodogram, knowledge about heteroscedastic behaviour, e.g. Poisson counting noise, is discarded and has to be inferred again from the periodogram.
These measurement uncertainties can be included in time-domain analyses and thus better inform the results.
Furthermore, using the time domain also obviates the need for signal windowing as there is no requirement for the time series to be of infinite duration.

Gaussian Processes (GPs; see \citet{Rasmussen2006} for an in depth introduction) describe a class of models that have enjoyed increasing popularity in astrophysics, owing to their flexibility and recent improvements in computational performance. They enable us to parametrize both long-term deterministic trends as well as stochastic processes in the same model via mean functions and covariance functions. While they have been used for a range of different astrophysical data sets, e.g.~as models for the cosmic microwave background~\citep{Bond1987, Bond1999, Wandelt2003}, they have been particularly popular in time series analysis.
For example, \citet{Moore2016} proposes to use GPs to account for uncertainty of gravitational-wave models and \citet{DEmilio2021} demonstrates how GPs can estimate the density of gravitational-wave posteriors.
Relevant to our goal here of finding quasi-periodicities, there is some work demonstrating the viability of using GPs to analyse or search for periodicities in light curves, such as blazars~\citep{Covino2020} and quasars~\citep{Zhu2020}.

The flexibility of the GP framework have made them a model of choice to account for structure in the data unimportant to the astrophysical problem of interest, e.g.~modeling correlated instrumental noise \citep{Gibson2012}, modeling stellar variability in transit exoplanet searches \citep{Grunblatt2015}, accounting for pulsar timing residuals \citep{VanHaasteren2014} or as a model for stellar spectra in radial velocity searches \citep{Czekala2017}. However, for certain problems, they can also be employed as empirical models that can be related to astrophysical properties even when the covariance function itself bears no direct connection to the underlying physics. Examples include as a model for stellar spots, where a quasi-periodic covariance function and a Gaussian Process can be used to infer stellar rotation \citep[e.g.][]{angus2018}, or in modeling stochastic variability in Active Galactic Nuclei \citep{Kelly2014}. For some problems, it is possible to construct covariance functions that are directly related to the underlying physical properties of the observed source, as was recently done for example by \citet{luger2021} in an extension of the earlier work on stellar spots, and by \citet{willeckelindberg2022} in modeling sparse asteroid light curves. 

In this paper we discuss the use of Gaussian Processes to perform searches for QPOs in the time domain. Focusing on transients specifically, we parametrize our model for the observed time series into a deterministic burst envelope, a stochastic component modeling rapid variability within the burst, and a candidate QPO, and simultaneously infer details of these components through a parametrization of each via distinct mean and covariance functions. 
Although creating and evaluating general non-stationary QPO models is complicated, we show how we can make some relatively easy tweaks to stationary models to explore non-stationary structures. 

Our approach here is situated somewhere between the second and third use case introduced above: as we will show below, the mean functions and covariance functions chosen to represent the bulk of the variability in all of our examples are strictly empirical and while they generally fit the data well, are not easily related to the astrophysical processes presumed to have generated the emission. On the other hand, the quasi-periodic oscillations we hope to find in transient light curves generally are directly related to astrophysical interpretations, including star quakes in magnetars and magnetic reconnection in solar flares.

We show that we can use Bayesian inference to determine the properties of a QPO and to reliably perform model selection between QPOs and alternative models such as red noise.
We lay out in detail what GPs are and motivate a kernel function that corresponds to QPOs in Sec.~\ref{sec:c6:gps}.
In Sec.~\ref{sec:injection_studies}, we demonstrate on simulated data that both parameter estimation and model selection work reliably.
Next, we demonstrate the viability on simulated light curves that \citet{Broomhall2019} produced for a comparative study between different QPO detection methods.
We show in Sec.~\ref{sec:real_data} that the method is easily transferable to many astrophysical data sets.
Specifically, we re-analyse GRB090709A~\citep{Cenko2010, DeLuca2010, Iwakiri2010}, a magnetar burst from SGR 0501~\citep{Rea2009, Huppenkothen2012}, and a segment from the SGR 1806-20 giant flare~\citep{Hurley2005, Palmer2005, Israel2005, Strohmayer2005, Strohmayer2006, Watts2006, Huppenkothen2014a, Huppenkothen2014b, Miller2019}.
Beyond the phenomena discussed in this paper, there have been recent claims about QPOs in fast radio bursts~\citep{Andersen2021, PastorMarazuela2022}, where the method introduced here provides an alternative approach to look for signals .
We discuss the results in Sec.~\ref{sec:discussion}.

\section{Methods}\label{sec:c6:gps}
In the following, we briefly introduce the core definitions of GPs, specifically in the one-dimensional case relevant to time series analysis.
For a more in-depth review, we refer the reader to~\citet{Rasmussen2006} and \citet{Foreman-Mackey2017}.

In the following, we distinguish \textit{deterministic} from \textit{stochastic} data generation processes. 
In short deterministic processes are those where the value of the time series at time $t_i$ can be exactly forecast given a mathematical description of the data generation process. 
For example, given an amplitude, a phase and a period, the value of a sinusoidal time series can be exactly forecast for any $t_i$. 
For stochastic processes, this is not true: here, the data generation process inherently involves a random element, such that future data points can only be described probabilistically. 
An example are random walk-type processes such as Brownian motion. It is relevant here to distinguish \textit{stationary} stochastic processes, where the mean and variance (weak stationarity) or the overall statistical properties (strong stationarity) of the process do not change with time, from \textit{non-stationary} stochastic processes, where the mean and variance and the statistical properties are time dependent. 
Stochastic processes are common in astronomy e.g.~in accreting sources. 
The wide variety of burst and flare shapes observed for example in magnetar bursts and GRBs easily lends itself to the assumption that the underlying process generating these bursts and flares is stochastic.

\subsection{Gaussian Processes Overview}
We can understand a GP as the sum of a deterministic process and a stochastic process.
The deterministic process takes $N$ time stamps $\{t_i\}_{i=1}^{N}$--which can be evenly or unevenly spaced--and maps them to a flux $y_i$.
The stochastic process assumes the data points are draws from a multivariate Gaussian distribution with dimensionality $N$, where the variances and covariances describe the temporal relationships between data points.
We can draw samples from the multivariate normal distribution similar to univariate distributions to use the GP as a generative model through a parametrization of the deterministic process--called the \textit{mean function}--and the stochastic process, generally described by a parametric model called the \textit{covariance function}. For forecasting purposes, the parameters of these functions are nuisance parameters. In the inference context of this paper, both mean function and covariance function have parameters are ascribed physical meaning, and thus their posterior distributions reveal interesting information about the system under study.

The covariance function $k_\alpha(t_i, t_j)$ with a vector of parameters\footnote{In the context of GP regression, the literature refers to the parameters $\alpha$ as ``hyperparameters", which should not be confused with the hyperparameters used in hierarchical inference.} $\alpha$ defines a covariance matrix $K(\alpha)$ of size $N\times N$. 
We write the mean function as $\mu_{\theta}(t)$, with parameter vector $\theta$.

We define the coordinate vector $\mathbf{t}$ (in our case, the coordinate of interest is time, but in practice, it need not be) and data vector $\mathbf{y}$ with $N$ entries each.
The GP log-likelihood is thus
\begin{equation}\label{eq:GP_likelihood}
    \ln L(\mathbf{\theta}, \mathbf{\alpha}) = -\frac{1}{2}\mathbf{r}_{\theta}^\mathbf{T}K_{\alpha}^{-1}\mathbf{r}_{\theta} - \frac{1}{2} \ln \mathrm{det} K_{\alpha} - \frac{N}{2} \ln(2\pi) \, ,
\end{equation}
where
\begin{equation}
        \mathbf{r_{\theta}} = \mathbf{y} - \mathbf{\mu_{\theta}} (\mathbf{t}) \, ,
\end{equation}
is the residual vector.
The covariance matrix elements are given by the kernel function $[K_{\alpha}]_{nm} = \sigma_n^2 \delta_{nm} + k_{\alpha}(t_n, t_m)$ where $\sigma_n^2$ are the variances due to white noise and $\delta_{nm}$ is the Kronecker delta.
In the simplest case of $k_{\alpha}(t_n, t_m) = 0$, the likelihood function becomes a simple Gaussian likelihood which one would typically employ in least-square regression.
The fact that we can include Gaussian measurement errors $\sigma_n$, which manifest themselves as a white noise floor in periodograms but can otherwise not be taken into account, means that we correctly treat heteroscedastic data.
In the above definition of the covariance matrix, we have assumed a stationary covariance function, such that the kernel only depends on the \textit{time difference} $\tau$ between different $y_i$, i.e. $k(t_n, t_m) = k(|t_n - t_m|) = k(\tau)$. We use different covariance functions to model broadband variability (also known as \textit{red noise}) as well as QPOs. Non-stationary covariance functions exist, but are computationally challenging to evaluate. For the majority of this work, we assume that a significant fraction of the variability in fast transients can accurately be described by a stationary stochastic process, and non-stationarity enters the model primarily through the mean function.

One of the main drawbacks of GPs is that, in general, the likelihood takes $\mathcal{O}(N^3)$ steps to evaluate~\citep{Rasmussen2006}.
There are numerous sophisticated approaches to reduce this complexity to a more manageable level~\citep{Wilson2015, Flaxman2015, Gardner2018}, including GPU acceleration~\citep{Delbridge2019}.
Most suitable for our purposes is the \celerite software package, which solves performance issues by restricting itself to one-dimensional problems and the class of sums and products of complex exponential kernel functions, which reduces the complexity to $\mathcal{O}(NJ^2)$~\citep{Foreman-Mackey2017} $J$ being the number of exponential terms that form the kernel.
\celerite achieves this by exploiting the semi-separable structure of the covariance matrices, which allows for the use of a fast solver for the Cholesky factorisation~\citep{Foreman-Mackey2017}.
We investigate in the following section why the class of exponential kernel functions is sufficient for our purposes.

\subsection{Kernel Functions}\label{sec:kernel_func}

\begin{figure*}
    \centering
    \includegraphics[width=\textwidth]{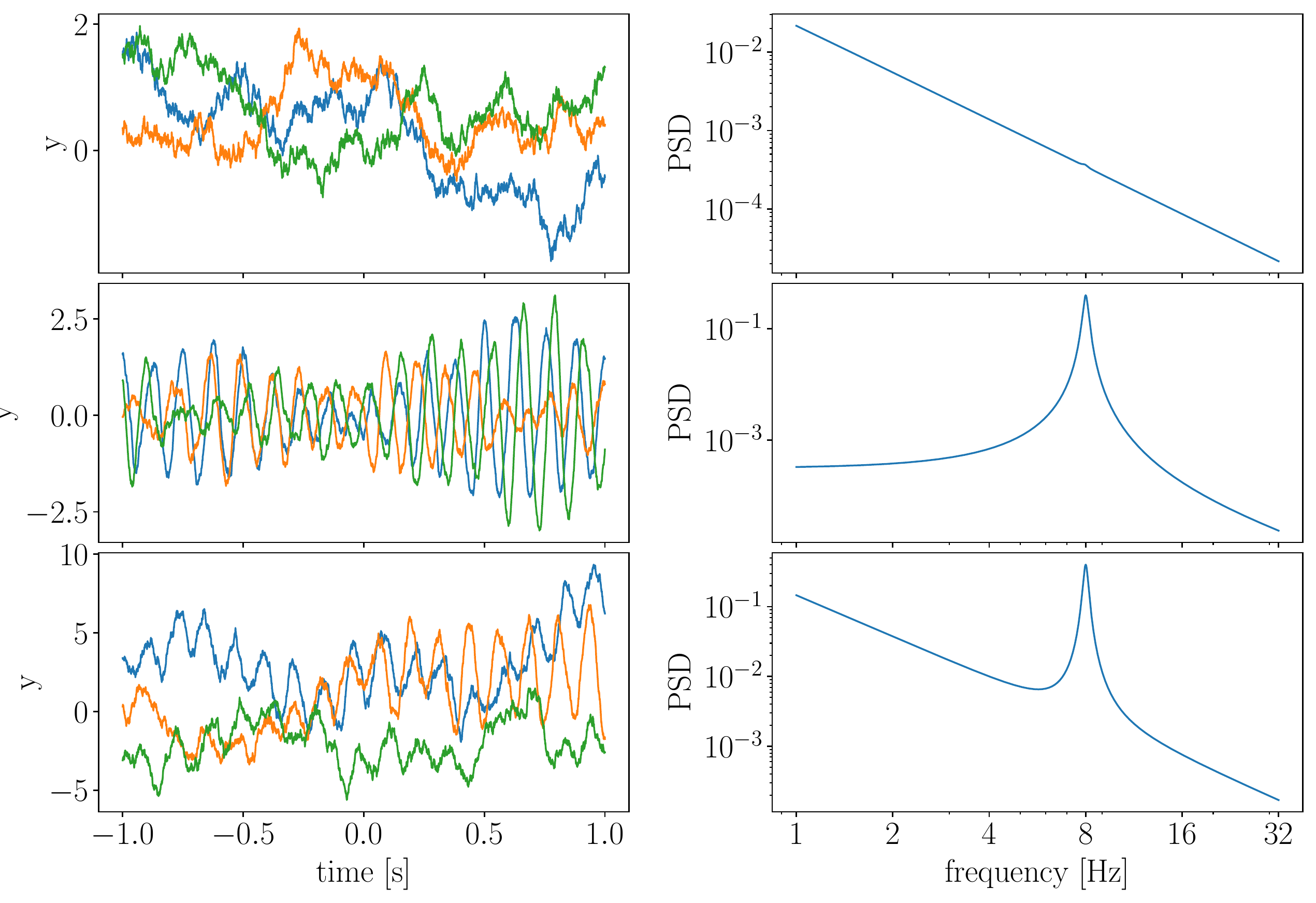}
    \caption{
    Realisations of GPs with a red noise $\krn$ (top left), quasi-periodic $\kqpo$ (middle left), and combined red noise and quasi-periodic $\krnqpo$ kernels (bottom left).
    The figures on the right are the corresponding power spectra to the time series on the left.
    We set $\aqpo = \arn = \cqpo = \crn = 1$ and $\fqpo = 8$ in the first and second panel, and use the same parameters except for $\arn = \exp (2)$ in the third panel.}
    \label{fig:realisations_gp}
\end{figure*}

There is an exhaustive literature on popular kernel functions, but it is not always clear which kernel function is most suitable to use and how to interpret their hyperparameters.
In fact, for many applications of GPs, this is not even the goal.
Instead, one may be interested in a GP's predictive power to interpolate and extrapolate data, which can be tested and tuned on training sets.
However, we are interested in interpreting the kernel function and associating it with an underlying physical process.
For example, we want to infer a posterior distribution of the frequency of a QPO.

We start by defining a kernel describing a periodic oscillation
\begin{equation}
    k_{\mathrm{po}}(\tau) = a \cos(2\pi f\tau) \, ,
\end{equation}
where $a$ is the amplitude of the oscillation and $f$ is its frequency.
This kernel corresponds to a perfect harmonic oscillation; its representation in the power spectrum is a Dirac delta function peaking at $f$.
In order to make this kernel quasi-periodic, we add another factor to account for variations in amplitude over time
\begin{align}\label{eq:qpo_kernel}
\kqpo(\tau) &= \exp(-c\tau) k_{\mathrm{po}}(\tau) \, ,
\end{align}
where $c$ is the inverse of the decay time of the QPO.
The $c$ parameter indicates that if we generated a time series using this kernel, the QPO would be excited or damped on a typical time scale of $1/c$.
In the frequency domain, this kernel corresponds to a Lorentzian function with peak frequency $f$ and full-width at half maximum $c$.
Since GPs are also generative models, we plot some realisations of the $\kqpo$ kernel in the upper left panel of Fig.~\ref{fig:realisations_gp}.

In practice, many astrophysical systems will display both red noise and QPOs at the same time.
In the context of \celerite, we can model red noise as a simple exponential  
\begin{equation}\label{eq:red_noise_kernel}
    \krn(\tau) = a \exp(-c\tau) \, .
\end{equation}
Physically, this kernel corresponds to a damped random walk (Ornstein-Uhlenbeck process), and its power spectrum is a $f^{-2}$ power law.
Since $\krn(\tau)$ is not mean-square differentiable, the model also implies that corresponding time series are not differentiable~\citep{Rasmussen2006}.
We show some realisations of $\krn$ in the upper left panel of Fig.~\ref{fig:realisations_gp}.

In order to obtain a kernel that describes the QPO and the red noise process, we simply add both kernels
\begin{align}\label{eq:qpo_plus_red_noise_kernel}
    \krnqpo(\tau) &= \kqpo(\tau) + \krn(\tau) \, .
\end{align}
This kernel is somewhat different to a QPO kernel proposed in \citet{Foreman-Mackey2017}
\begin{equation}
    k(\tau) = \frac{a}{2 + b} \exp(-c\tau) \left[\cos(2\pi f\tau) + (1 + b) \right]\, ,
\end{equation}
which also both models QPO and red noise features, but makes them share the $c$ parameter.
Since we do not find that $c_{\mathrm{qpo}} = c_{\mathrm{rn}}$ in general, we use Eq.~\ref{eq:qpo_plus_red_noise_kernel} as our model for the general case and Eq.~\ref{eq:red_noise_kernel} for the case that we only encounter red noise.

There are alternative ways to describe periodic behaviour outside the domain of \celerite compatible kernel functions.
\citet{Rasmussen2006} used the following kernel to model seasonal changes in atmospheric CO2 levels
\begin{equation}\label{eq:Keeling_kernel}
    k(\tau ) = a\exp\left[ -\frac{\tau^2}{2\ell^2} - \Gamma \sin^2 \left( \pi f \tau \right) \right] \, .
\end{equation}
\citet{angus2018} demonstrated that this kernel can be used to infer stellar rotation periods.
This kernel combines a squared exponential with an oscillatory term such that a parameter $\Gamma$ controls the amount of covariance between two points that are roughly one period away from each other.
For high values of $\Gamma$, only points exactly integer multiples of a period away from each other have high covariance.
Similarly to $\kqpo$, Eq.~\ref{eq:Keeling_kernel} is quasi-periodic in that it is damped, though not exponentially but with the square-exponential.
This way, arbitrary curves with a repeating shape can be modelled, though a QPO kernel as in Eq.~\ref{eq:qpo_plus_red_noise_kernel} has shown to be sufficient to correctly infer the frequency for arbitrary oscillatory curves~\citep{Foreman-Mackey2017}.
Furthermore, while Eq.~\ref{eq:Keeling_kernel} is practical to fit non-sinusoidal curves, it may be hard to interpret the underlying physics from the inferred parameters and comes with the high computational cost of general GP kernels.
Alternatively, Eq.~\ref{eq:Keeling_kernel} could also be approximated by a Fourier series since the \celerite kernel family forms a Fourier basis.
However, this approach would also incur far higher computational costs since the \celerite likelihood complexity scales quadratically with the number of terms required to build the kernel, and the additional Fourier coefficients would drastically increase the parameter space.

There are also alternative ways to define aperiodic kernels.
Within the support of \celerite models, \citet{Foreman-Mackey2017} proposed using a critically damped stochastic harmonic oscillator
\begin{equation}
    k(\tau ) = S_0 \omega_0 e^{-\frac{1}{\sqrt{2}}\omega_0 \tau} \cos \left( \frac{\omega_0 \tau}{\sqrt{2} - \frac{\pi}{4}}\right)\, ,
\end{equation}
which is commonly used to model background granulation noise in asteroseismic and helioseismic oscillations.
The stochastic harmonic oscillator is once mean-square differentiable and thus yields smoother time series data than the red noise kernel.
Outside of the class of sums and products of complex exponentials that \celerite supports, the squared exponential, Matern-3/2 and rational square kernel functions are employed in many contexts.
Recently, some of these kernels have also been applied to search for oscillations in blazar light curves~\citep{Covino2020}. 
However, these kernels are difficult to motivate physically, and the higher computational complextity may restrict their use to maximum likelihood estimates rather than Bayesian evidence calculations.
We stick to the red noise kernel for our analyses as we consider this to be a basic framework that may describe many physical systems to a first degree.

\begin{figure}[h]
    \centering
    \includegraphics[width=\columnwidth]{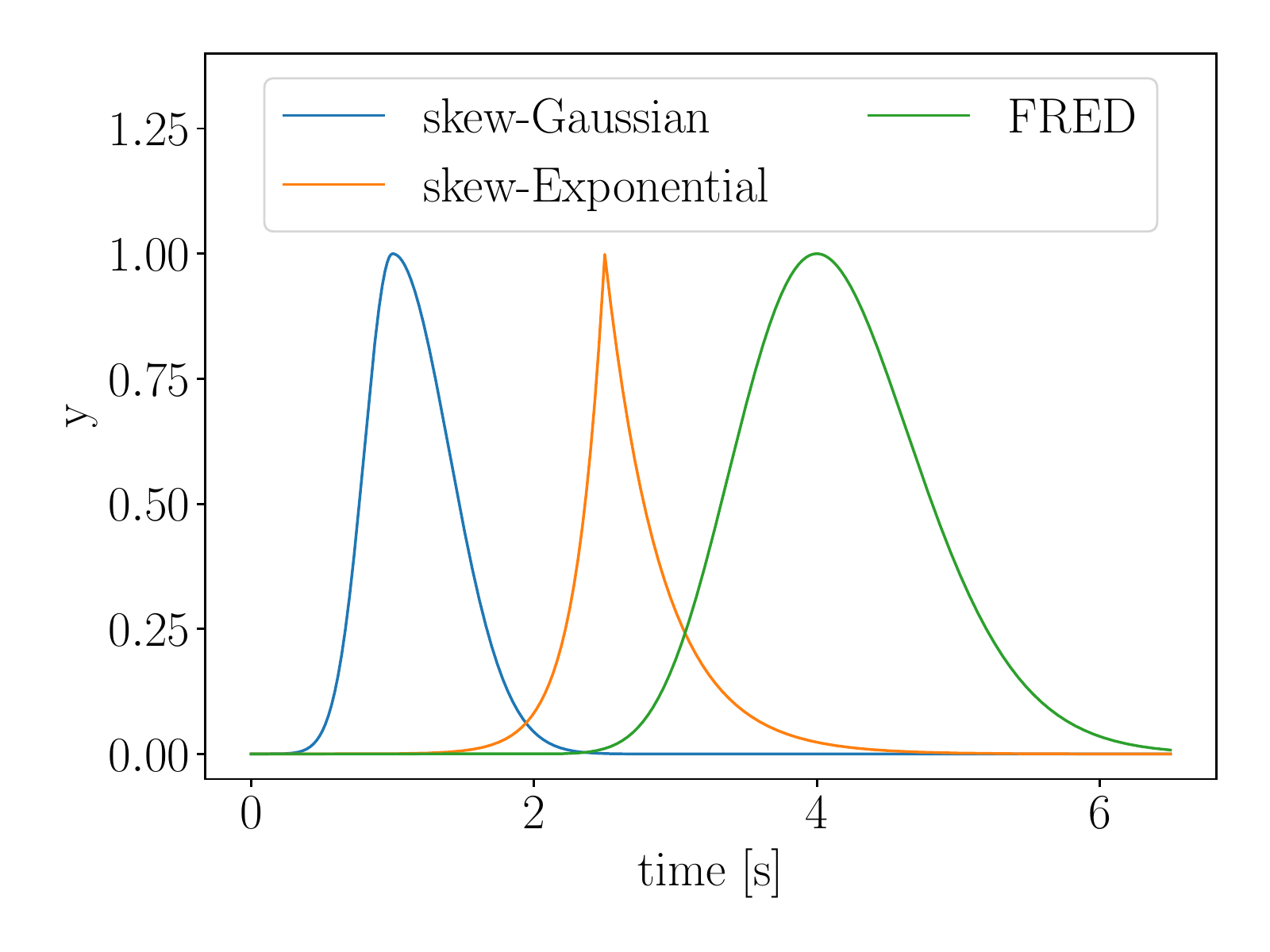}
    \caption{Example plots of the three mean models we use throughout.}
\label{fig:realisations_mean}
\end{figure}
\subsection{Mean Functions}\label{sec:mean_functions}

Choosing a mean model is similarly tricky to choosing a kernel function if there is no physically-motivated model for the underlying background trend.
The simplest mean model takes the (weighted) average of all recorded data points.
However, this model implicitly assumes that there are no trends in the data and thus all variability in the light curve was generated by a stochastic process that can be described by the chosen covariance function.
If a significant background trend is present, this assumption quickly breaks down and leads to incorrect inferences about the variability in the light curve.
It is thus not a suitable way to approach the problem in most cases.

Another approach is to use filter methods such as the boxcar filter or the Savitzky-Golay filter~\citep{Savitzky1964} and subtract the filtered light curve from the original light curve.
However, using filtering methods is prone to cause artificial periodicities~\citep{Auchere2016}.
Some problems feature a relatively simple background trend that one could model as a linear trend or another low-order polynomial instead of a filter.
For some problems, physical or phenomenological models may exist to describe the overall trend of the light curve, such as the ``fast-rise exponential decay'' (FRED) model for GRB light curves.

For general problems, one can employ methods such as shapelet fitting, in which one casts the problem as a sum of orthonormal basis functions.
Alternatively, for burst-like light curves, it is often possible to identify several base flare shapes (e.g. (skewed) Gaussians, FREDs; \citealt{Huppenkothen2015}) that one can use to model the overall shape of the burst.

Here, we use a number of very simple phenomenological prescriptions as flare mean models, and urge the reader to consider physically meaningful options for mean functions in their specific application.
As the most basic models we use skewed Gaussians and skewed exponentials, i.e.
\begin{equation}
    \mu_{\mathrm{GAUSS}}(t; A, t_0, \sigma_1, \sigma_2) = 
    \begin{cases}
        A \exp \left(- \frac{(t - t_0)^2}{2\sigma_1^2} \right)\, \mathrm{if}\, t \leq t_0
        \\
        A \exp \left(- \frac{(t - t_0)^2}{2\sigma_2^2} \right)\, \mathrm{if}\, t > t_0
        \end{cases}
\end{equation}
and
\begin{equation}
    \mu_{\mathrm{EXP}}(t; A, t_0, \sigma_1, \sigma_2) = 
    \begin{cases}
        A \exp \left(- \frac{t - t_0}{\sigma_1} \right)\, \mathrm{if}\, t \leq t_0
        \\
        A \exp \left(\frac{t - t_0}{\sigma_2} \right)\, \mathrm{if}\, t > t_0 \, .
        \end{cases}
\end{equation}
There are a number of other approaches to define FRED models. 
We focus on the FRED model defined in \citet{Norris1996}, which can assume a wide variety of shapes.
\begin{align}\label{eq:fred}
\begin{split}
    \mu_{\mathrm{FRED}}(t; A, t_0, &\psi, \Delta) =\\ &A\exp\left[ -\psi\left(\frac{t + \Delta}{t_0} + \frac{t_0}{t + \Delta} \right)\right] \exp(2\psi)
\end{split}
\end{align}
for $t < 0$.
This model's maximum coincides with $t = t_0$, and we can interpret $\Delta$ as an ``offset" parameter that allows us to fit the full range of flare shapes with arbitrary offsets from $t = 0$.
Finally, $\psi$ serves as a symmetry parameter, with the curve becoming increasingly symmetrical for large values of $\psi$.
The additional factor $\exp(2\psi)$ is optional in the model definition and serves as a normalisation factor such that $\mu(t=t_0; A, t_0, \psi, 0) = A$, which eases the definition of the amplitude prior.

We display the mean models in Fig.~\ref{fig:realisations_mean}.
As we show later, all models we have introduced here reasonably fit general flare shapes.
Performing Bayesian inference with these models thus naturally yields comparable evidences. In this paper, we consider models that contain a single instance of one of these mean models, but we also consider mean functions that are composed of a superposition of multiple components, e.g.~a model with three three skew-Gaussian components, each with their own parameters. 
For general problems, it is prudent to run the analysis with multiple mean models and select the one that yields the highest evidence.

\subsection{Towards non-stationary models}\label{sec:c6:towards_non_stat_models}
We have so far laid out how GPs resolve the issues with heteroscedastic uncertainties and deterministic trends that lead to biases in Fourier-based analyses.
The other issue we face is if the QPO and noise process are non-stationary. Bursts and flares are inherently non-stationary processes: it is possible that the properties of the noise process change as a function of time within a burst. Similarly, it is possible that the QPO is only present for part of the burst, as is the case in neutron star burst oscillations\citep{Watts2012a}.

If we do not select the data segment we want to analyse carefully, we are likely to incur similar non-stationarity biases as periodogram-based analyses as \citet{Huebner2021} laid out. However, we do not know \textit{a priori} whether the variability processes in the data are stationary or not, and whether a QPO may exist over the full length of the light curve or only part of it. Ideally, we would like to infer these properties, rather than assume stationarity over the full time series.

As a simple example, we consider a system with constant X-ray emission characterised by white noise, which flares for some time, and then goes back to its regular constant emission.
We describe this scenario by first defining a continuous mean function to be present through the entire time series (here a constant, though in practice, it could be more complex). 
Additional variability on top of that shape is present in three distinct segments, where the stochastic variability in the first and last segments are described solely by white noise, and the middle segment which also contains both a QPO and red noise.
We show how we can implement this basic, non-stationary process and use it to infer the QPO's start and end.

By breaking the time series into three segments, we can model a basic, non-stationary process without resorting to a more general likelihood solver.
Instead, we split the likelihood into two parts, one for the two disjoint, white noise segments on either end and the GP in the centre.
The covariance matrix can then be decomposed into two submatrices correspondingly.
We obtain one ``outer'' submatrix by deleting all rows and columns that contain off-diagonal entries, and one covariant ``inner'' submatrix as the complement of the outer submatrix.
Since the inner covariant submatrix forms a block, we can calculate the inverse and the determinant independent from the remaining entries.
The same operations are trivial for the outer submatrix. 
We re-write Eq.~\ref{eq:GP_likelihood} to describe this explicitly
\begin{align}\label{eq:GP_likelihood_windowed}
\begin{split}
    \ln L(\mathbf{\theta}, \mathbf{\alpha}) 
    = &-\frac{1}{2} \sum_{n_{\mathrm{out}}} \frac{r_{n_{\mathrm{out}}}^2}{\sigma_n^2} - \sum_{n_{\mathrm{out}}} \ln\sigma_{n_{\mathrm{out}}} - \frac{N_{\mathrm{out}}}{2} \ln(2\pi) \\
    &-\frac{1}{2}r_{\mathrm{in}, \theta}^\mathbf{T}K_{\mathrm{in}, \alpha}^{-1}r_{\mathrm{in}, \theta} - \frac{1}{2} \ln \mathrm{det} K_{\mathrm{in}, \alpha} - \frac{N_{\mathrm{in}}}{2} \ln(2\pi) \, 
\end{split}
\end{align}
where the likelihood coming from the outer non-covariant submatrix is in the first line and the inner covariant submatrix is in the second line.
Here $n_{\mathrm{out}}$ are the indices of the outer $N_{\mathrm{out}}$ non-covariant row/columns, $K_{\mathrm{in}, \alpha}$ is the inner covariant submatrix, and $r_{\mathrm{in}}$ are the residuals of the $N_{\mathrm{in}}$ inner rows/columns.
We introduce the $t_{\mathrm{start}}$ and $t_{\mathrm{end}}$ parameters to our model to describe the transition points between the segments, to be inferred together with the remaining parameters.

\subsection{Assumptions and Limits of GPs}\label{sec:limits}
As with all other methods, the reliable detection of QPOs in time series crucially depends on whether the assumptions of the method and of the model proposed to generate the data are reasonable ones. 
For many of the examples we showcase in this paper, no realistic physical model exists capable of generating realistic simulated data. 
Therefore, QPO detection necessarily relies on imperfect, phenomenological prescriptions to describe the data generation process--with or without QPO. 
Here, we assume that the observed time series can be described as the combination of a simple flare shape and a stochastic process. 
This is a more realistic extension of the assumptions that all variability can be described by a stationary stochastic process that previous approaches using Fourier methods have made. 
As our understanding of these phenomena improves, better assumptions about the physical processes that generates these bursts and flares will significantly improve QPO detections.

One fundamental assumption we make is that QPOs are an additive process, i.e. the QPO does not depend on the value of the mean function at any point.
This assumption is likely not a perfect model.
At the very least, we expect no QPOs long before and after a transient.
This observation implies some connection between the mean function and the GP, which is challenging to model using our framework.

The QPO model we propose is also inherently stationary as this is required for the fast Cholesky solver implemented in the \celerite package, and we can only model a narrowly defined set of non-stationary time series as we have shown in Sec.~\ref{sec:c6:towards_non_stat_models}.
Quasi-periodic oscillations that drift in frequency over long time scales can not be adequately modelled and require us to use a different framework of GP solvers.
Once frequency shifts become too large, our QPO kernel is sure to fail.
We can detect shifting frequencies by splitting the light curve into multiple segments that we analyse independently and compare the QPO frequency posteriors, or via hierarchical Bayesian inference.
We would have to evaluate a more expensive non-stationary QPO model for a more rigorous analysis.

Many of the phenomena considered here are bright, energetic flares observed in X-rays. Observatories typically record astrophysical X-ray data by counting the number of photons that arrive at a detector in a specific time interval, a process by its nature Poissonian. GP modelling however inherently assumes that the underlying data are Gaussian.
While Poisson counting data are Gaussian to a good approximation if there are sufficient counts per bin, GPs do not correctly model data with lower count rates.
There are some approaches to deal with photon-counting data with low count rates.
The easiest way is to apply a variance stabilising transform, which makes the data approximately Gaussian with $\sigma = 1$~\citep{Anscombe1948, Bar-Lev1990}.
However, these transformations are still far off for bins with zero or one photon.
Other methods like a sigmoidal Gaussian-Cox process treat Poisson data effectively as Gaussian by converting the data with a sigmoid function~\citep{Adams2009, Flaxman2015}, but using these methods makes it harder to interpret the parameters because the sigmoid function also transforms them.

Real data is often more complicated than a simple Poisson process and may feature additional effects such as dead time or double-counting of photons.
We assume that dead time, i.e. the property of detectors that they cannot detect a second photon for a short period after they counted the first photon, does not play a significant role in our data.
Dead-time effects are particularly a problem for strongly flaring sources.
Modelling dead-time in practice is complex and requires sophisticated techniques such as simulation-based inference~\citep{Huppenkothen2021}.

\subsection{Bayesian Inference with Gaussian Processes}\label{sec:bay_inf_w_GPs}
Bayesian inference is a statistical paradigm for parameter estimation and model selection. 
We used Bayesian inference with the calculation of Bayes factors in \citet{Huebner2021} to characterize model preference and thereby the significance of QPOs within time series data, and apply the same approach here.
To understand the meaning of Bayes factors, we consider Bayes' theorem
\begin{equation}
    p(\theta|d, M) = \frac{\pi(\theta|M)L(d|\theta, M)}{Z(d| M)} \, ,
\end{equation}
where $\theta$ are both the mean and kernel parameters, $d$ are the data, i.e. the time series in our case, $p(\theta|d, M)$ is the posterior probability of the parameters, $\pi(\theta|M)$ is the prior probability of the parameters, $L(d|\theta, M)$ is the likelihood of the data given the parameters, and $Z(d| M)$ is the evidence, or fully marginalized likelihood.
All these probabilities are conditioned on the combined kernel and mean model $M$, which we want to evaluate.
The evidence describes an overall probability of the given model producing the data, and can be calculated by rearranging and integrating Bayes' theorem
\begin{equation}\label{eq:evidence}
    Z(d| S) = \int \pi(\theta)L(d|\theta, S) d\theta \, .
\end{equation}
Equation~\ref{eq:evidence} is expensive to solve using simple grid methods, which is why we have to employ sophisticated algorithms such as nested sampling in practice~\citep{Skilling2004, Skilling2006}.

Though the evidence itself carries no straightforward intrinsic meaning as a normalization factor, taking the ratio between two yields the Bayes factor
\begin{equation}
    BF = \frac{Z(d| M_1)}{Z(d| M_2)}\, ,
\end{equation}
where $M_1$ and $M_2$ are the different models that yield the respective evidences. 
The Bayes factor measures the odds of the underlying data being produced by either model, assuming both models are equally likely to be correct, though it does not measure if the model itself is a good fit to the data. Because Bayes factors are sensitive to the prior volume, care should be taken when choosing priors and interpreting the results.

Throughout this paper we use the \bilby package to implement Bayesian inference~\citep{Ashton2018, Romero-Shaw2020}.
We wrote a small likelihood interface for \celerite and \bilby, which is now part of the main \bilby package.
All software used to produce the results in this paper can be found in its repository\footnote{\href{https://github.com/MoritzThomasHuebner/QPOEstimation}{https://github.com/MoritzThomasHuebner/QPOEstimation}}.

Bayes factor calculations for model selection of GPs usually have not been considered viable (\citet{Covino2020, Zhu2020} are notable exceptions).
One can compute a maximum likelihood estimate much faster and derive alternative measures, such as the Akaike or Bayesian information criterion.
However, these measures are less reliable if the posterior distribution is not Gaussian or features multiple modes.
When dealing with \celerite models, performing the necessary likelihood evaluations for complete posterior and evidence calculations is relatively fast.
We find that nested sampling can be performed in $\mathcal{O}(\mathrm{minutes})$ using \celerite for a light curve with a few hundred $t_i$ when using the $\kqpo$ GP kernel and a skewed Gaussian mean model.
Furthermore, even beyond the class of \celerite models, full Bayesian inference may be possible on many data sets with the use of efficient GP evaluation algorithms~\citep{Flaxman2015, Wilson2015, Gardner2018, Delbridge2019} or the use of massively parallel inference~\citep{Smith2020}.

\subsection{Priors}\label{sec:priors}
We must choose priors for the parameters of the covariance function $\alpha$ and of the mean function $\theta$ carefully. Here, we largely choose uninformative priors not tied to a specific application, but we urge readers to consider informative priors for their specific application carefully. Making priors informative is important, because the prior range impacts the evidence calculation: an overly large prior range would suppress the evidence and disfavour the model. However, in problems where knowledge about the underlying physical processes is sparse, setting informed priors is often not possible or impractical. For example, the possible amplitudes $a$ depend on both the efficiency of the instrument and the intrinsic brightness of the object we observe. Thus, we set prior bounds for amplitudes based on the data.

Generally, we employ priors uniform in logarithm for scale parameters of the covariance function such as the amplitude $a$, inverse decay time $c$, and frequency $f$ since these prior spaces span several orders of magnitude.

In theory, we can set the prior range of $\arn$ and $\aqpo$ arbitrarily wide.
However, in some circumstances this prior choice causes numerical issues.
For example, if there is no QPO present in the data, the $\aqpo$ posterior will have an upper limit below which it is identical to the prior distribution. 
Thus, we set the lower limit of both the $\arn$ and $\aqpo$ prior to be the smallest $\sigma_n$, meaning that we assume that the QPO/red noise process has to be at least as strong as the white noise.
We set the upper limit on $\arn$ and $\aqpo$ to be twice the difference between the highest and lowest value within the time series.

Assuming the data points are equally spaced, the frequency parameter $f$ should not be smaller than than $1/T$, with $T$ being the length of the data segment and not being greater than the Nyquist frequency, which is half the sampling rate.
The Nyquist frequency for unevenly sampled data is $1/2p$, where $p$ is the largest value such that each time value $t_i$ can be written as $t_i = t_0 + n_ip$ with $n_i$ being an integer~\citep{Eyer1999, VanderPlas2018}.
If one considers a specific source with some frequency range of interest, it may be useful to restrict the prior further.

The upper limit of $\crn$ should be not greater than the sampling frequency $f_s = 1/\Delta t$ of the data because otherwise, the kernel will decay to less than $1/e$ within $\Delta t$. 
Beyond this point, the red noise model is not functionally different from a white noise model.
If we consider unevenly sampled data, we can use the twice Nyquist frequency as an upper limit to reflect the choice for $f$.
We set the lower limit of $\crn$ to $1/T$ to ensure that the red noise process occurs on a time scale at a maximum still comparable to $T$.
While lower values of $\crn$ may have posterior support if permitted, they imply a constant offset between the mean function and the data that is not physically motivated.

Unlike with $\crn$, a very low $\cqpo$ cannot cause a constant offset since the $\kqpo$ averages to zero over long time scales.
Thus, $\cqpo$ can, in principle, take on infinitely low values.
We set the minimum of $\cqpo$ to be $1/(10T)$, which means that a $\kqpo$ has to decrease by $\approx 10\%$ throughout the data segment as we cannot functionally distinguish lower values.
For the upper limit of $\cqpo$ we need to ensure that $\kqpo$ is periodic.
Thus, we demand that $\kqpo$ does not decay to less than $1/e$ over one period, which implies $\cqpo \leq \fqpo$.
We implement this relation in practice using the \texttt{Constraint} prior feature in \bilby.

Finally, we also consider the priors of our analytic mean flare shapes.
Again, these generally depend on the details of the underlying physical process and our knowledge of it. For the bursts and flares considered in this paper, many of our priors are derived directly from properties of the observations, under the assumption that the properties of transient detection are often driven by the source's behaviour. For example, under the assumption that the instrument observed the entire flare down to some background, we can choose a uniform prior for the parameter describing the peak of the mean function limited to the start and end of the observation.
The amplitude priors are uniform in logarithm between 10\% and 200\% of the difference between minimum and maximum of the data.
The prior on the width parameter of our flares depends on the model function used for the mean model.
In general, we aim for the width parameter to have a uniform-in-logarithm prior that spans roughly from the time difference between two data points to the entire time range of the data.

For mean models defined as a superposition of multiple simple shapes, the prior for the peak time requires careful thought because the parameters of each model component are degenerate under permutation of the order of these components.
If we did not break this degeneracy, the parameter would be much harder to sample since $n$ mean model components allow for $n!$ degenerate permutations.
Assuming $n$ components making up the mean model throughout the interval, we can use a result from order statistics to derive appropriate priors.
We label the flares in order from $0$ to $n - 1$ indexed by $k$.
The $t_{0, k}$ prior is distributed according to a conditional beta distribution with $\alpha=1$ and $\beta= n - k$ between the $t_{0, k-1}$ flare and the end of the interval~\citep{Gentle2009}.
Putting this together, we obtain
\begin{equation}\label{eq:conditional_beta}
    \pi(t_{0, k}| t_{0, k-1}) =  (n-k) \left(1-\frac{t_{0, k}-t_{0, k-1}}{t_{\mathrm{max}}-t_{0, k-1}} \right)^{n-k-1} \, ,
\end{equation}
where $t_{\mathrm{max}}$ is the time of the last element of the light curve.
For the special case of $k = 0$ the prior is not conditional and we have instead
\begin{equation}\label{eq:minimum_prior}
    \pi(t_{0, k=0}) =  n \left(1-\frac{t_{0, k=0}-t_{\mathrm{min}}}{t_{\mathrm{max}}-t_{\mathrm{min}}} \right)^{n-1} \, ,
\end{equation}
where $t_{\mathrm{min}}$ is the first element of the light curve.

The priors on the parameters of non-stationary extension of the \celerite model, $t_{\mathrm{start}}$, and $t_{\mathrm{end}}$, are chosen in the same way as in Eq.~\ref{eq:conditional_beta} and Eq.~\ref{eq:minimum_prior}, but with $n=2$.

As a final note, the priors here are chosen to be empirical and relatively uninformative, with the goal of showcasing the procedure rather than enabling astrophysical inference on a specific problem. Careful design of the priors for any given physics problem is important in the context of \textit{population inference}, where the priors described above would likely lead to significant biases.

\section{Simulated Data}\label{sec:injection_studies}
In this section, we look at different configurations of the models we described earlier and validate them on simulated observations.
We examine the QPO kernel and our red noise kernel with different parameters.
We also investigate what happens when we infer parameters with the wrong model, e.g. if we create red noise data and perform inference with $\krnqpo$.

We generate GP data using a set of parameters in two steps.
First, we obtain the deterministic part of the data by evaluating the mean function.
Second, we produce the stochastic process arising from the covariance between the data points.
We create the stochastic process data by drawing a sample from the multivariate normal distribution associated with this covariance matrix.
Finally, we add the deterministic and stochastic parts to create the overall time series. 

\subsection{Percentile-Percentile Analysis}\label{sec:pp_analysis}
Before we begin applying our methods, it is important to establish their validity on simulated data sets.
Specifically, it is hard to know a priori how to tune sampling settings within \bilby and the nested sampling package we use here, \dynesty \citep{speagle2020}, to obtain adequate results.
Though gravitational-wave researchers frequently use \bilby and have established settings~\citep{Romero-Shaw2020}, the data sets and models we use have different dimensionality and posterior distributions that may be more difficult to sample from.
Specifically, we want establish that the posterior samples we draw are representative of the true posterior distribution.

If the inference process is unbiased, then the actual value of the parameters has to be in the $x$th-percentile of the posterior distribution $x$ per cent of the time.
We can use this percentile-percentile (PP) analysis to tune the sampling settings based on simulated data before deploying the analysis on real data~\citep{Cook2006}.
The PP analysis is a particular case of a quantile-quantile analysis in that both considered distributions are uniform between 0 and 1.
Percentile-percentile tests fail if the settings are insufficient to sample the posterior without bias, e.g. if we use too few live points or if some of the parameters are hard to sample.
Aside from difficulties in sampling, PP tests should pass when using the same model and priors for data creation and inference.
Concretely, the PP test finds a p-value for each parameter using a Kolmogorov–Smirnov test~\citep{Massey1951} to check if the fraction of events in a particular credible interval is drawn from a uniform distribution.
The PP test combines these individual p-values to a combined p-value.
Conventionally, $p < 0.05$ indicates that the fraction of events is not uniformly distributed to the $2\sigma$-level.

\begin{figure}
    \centering
    \includegraphics[width=\columnwidth]{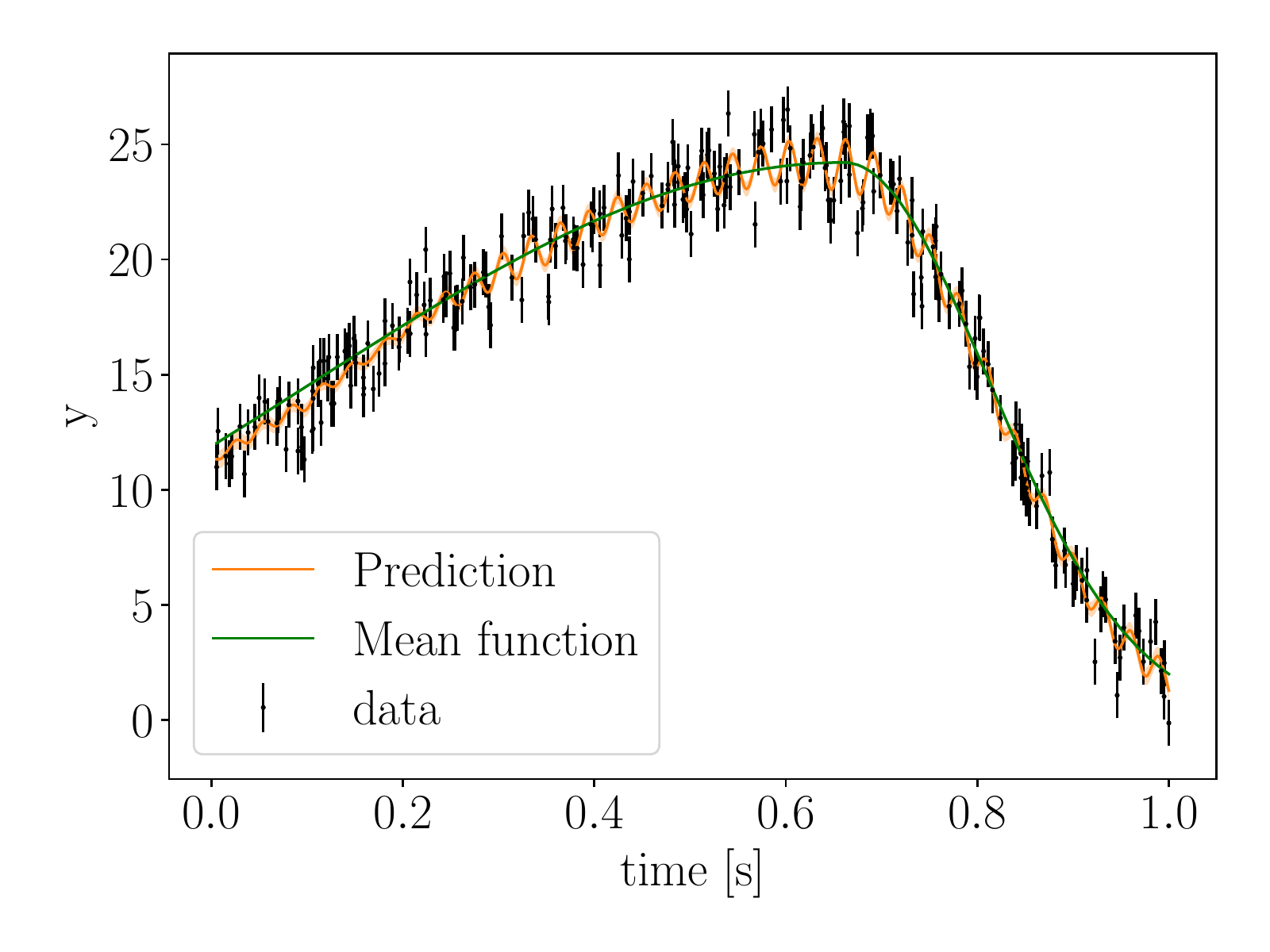}
    \caption{Example draw from the prior distributions given by the parameters in Tab.~\ref{tab:pp_injection_priors}. 
    We show the data in black and the mean function in green.
    The prediction curve with $68\%$confidence bands in orange shows, based on the underlying parameters, is an estimate of the $y$-values between the data points.}
    \label{fig:pp_injection_example}
\end{figure}

We perform the PP test using the $\krnqpo$ model and a single skewed Gaussian as an example mean model.
We create 100 simulated \SI{1}{\second} long data sets sampled with 256 time bins $t_{i}$ that are randomly uniform distributed in time and using parameters randomly drawn from the priors in Tab.~\ref{tab:pp_injection_priors}.
We display one random draw in Fig.~\ref{fig:pp_injection_example}.
Choosing randomly spaced $t_i$ does not require any additional effort since the prior boundaries are fixed and has the benefit of validating the method for a broader set of possible problems.

\begin{figure}
    \centering
    \includegraphics[width=\columnwidth]{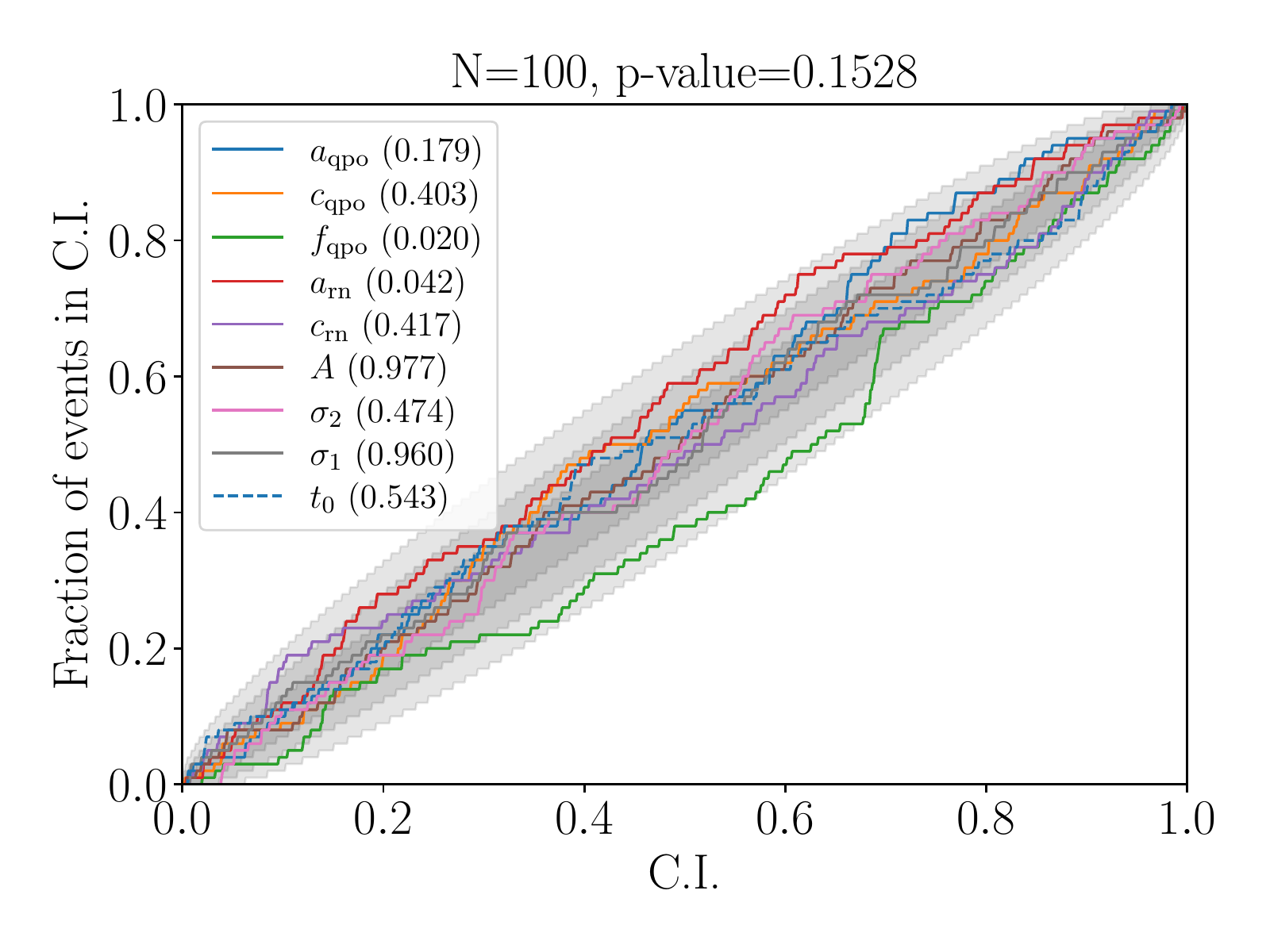}
    \caption
    {Percentile-percentile plot shows that our sampling methods are not biased.
    We obtain the plot using the respective function in \bilby~\citep{Romero-Shaw2020}. 
    We show the confidence interval (CI) on the horizontal axis and what fraction of events have the true value within the CI on the vertical axis.
    The grey bands indicate the $1\sigma$, $2\sigma$, and $3\sigma$-levels.
    The overall p-value of $0.1528$ indicates that the observed deviations are consistent with randomness, though the individual relatively low values for $\fqpo$ and $\arn$ shown in the parentheses in the legend indicate that these parameters may be harder to sample and may benefit from more finely tuned settings.}
    \label{fig:pp_plot}
\end{figure}

Figure~\ref{fig:pp_plot} shows the PP analysis using \dynesty's random walk sampling with 1500 live points and \bilby default settings.
This result shows that these setting are sufficient to sample the parameter space without significant biases in the posterior.
The overall p-value of $0.1528$ indicates that the observed deviations are consistent with randomness, though the individual relatively low values for $\fqpo$ and $\arn$ indicate that these parameters may be harder to sample and may benefit from more finely tuned settings.
We find similar results using \dynesty's random slice sampling, which is substantially faster but generally produces a worse representation of the posterior~\citep{Romero-Shaw2020}.

\begin{table}[ht]
    \centering
    \begin{tabular}{c|c|c|c}
        Parameter   & Prior class           & Minimum   & Maximum \\ \hline\hline
        $\arn$      & \texttt{LogUniform}   & $\exp(-1)$& $\exp(1)$\\ \hline
        $\crn$      & \texttt{LogUniform}   & $\exp(-1)\, \si{\per\second}$& $\exp(1)\, \si{\per\second}$\\ \hline
        $\aqpo$     & \texttt{LogUniform}   & $\exp(-1)$& $\exp(1)$\\ \hline
        $\cqpo$     & \texttt{LogUniform}   & $\exp(-1)\, \si{\per\second}$& $\exp(1)\, \si{\per\second}$\\ \hline
        $\fqpo$     & \texttt{LogUniform}   & $\SI{1}{\hertz}$ & $\SI{64}{\hertz}$\\ \hline
        $A$         & \texttt{LogUniform}   & 10 & 100\\ \hline
        $t_0$       & \texttt{Uniform}      & $\SI{0}{\second}$ & $\SI{1}{\second}$\\ \hline
        $\sigma_1$  & \texttt{LogUniform}   & $\SI{0.1}{\second}$ & $\SI{1}{\second}$\\ \hline
        $\sigma_2$  & \texttt{LogUniform}   & $\SI{0.1}{\second}$& $\SI{1}{\second}$\\ 
    \end{tabular}
    \caption{Priors for PP test. The prior is used to randomly draw parameters for simulated data sets and during the inference process.
    There is also a prior constraint such that $\cqpo < \fqpo$, which we introduced in Sec.~\ref{sec:bay_inf_w_GPs}.}
    \label{tab:pp_injection_priors}
\end{table}

\subsection{Model Selection}\label{sec:model_selection}
We want to understand how stable and reliable Bayes factors are in QPO detection given multiple light curves generated from a model with the same parameters, since noise realizations of the same parameters can be vastly different.
For this study, we create the data with a symmetric Gaussian mean function.
We create one set of simulated light curves using $\krn$, and two sets using $\krnqpo$ with different QPO amplitudes.
We list the parameters in Tab.~\ref{tab:mss_injection_priors}.
The first QPO amplitude is relatively low compared to the noise in the data, whereas the higher QPO amplitude corresponds to a four-fold increase over the lower one.
We use both kernel functions plus the skewed Gaussian mean model to carry out model selection between a model with a QPO and a model without a QPO for all light curves. 
We define the QPO Bayes factor
\begin{equation}
    \BFQPO = \frac{Z(d|\krnqpo, \mu)}{Z(d|\krn, \mu)} \, ,
\end{equation}
where the numerator and denominator are the respective evidences given either kernel function and the same mean model $\mu$.
We expect that $\ln \BFQPO$ should generally be positive if we created the data using $\krnqpo$ and negative otherwise.
We create the data on a \SI{1}{\second} interval sampled equidistantly at \SI{256}{\hertz} using the parameters listed in Tab.~\ref{tab:mss_injection_priors}.
We choose equidistant sampling in this case because prior choices do matter for model selection and we want to validate model selection with the prior ranges we define in Sec.~\ref{sec:bay_inf_w_GPs} and apply on real data.
Specifically, the Nyquist frequency for unevenly sampled data can be orders of magnitude greater than for evenly sampled data, thus greatly enlarging the prior volume of $\fqpo$.

\begin{figure}
    \centering
    \includegraphics[width=\columnwidth]{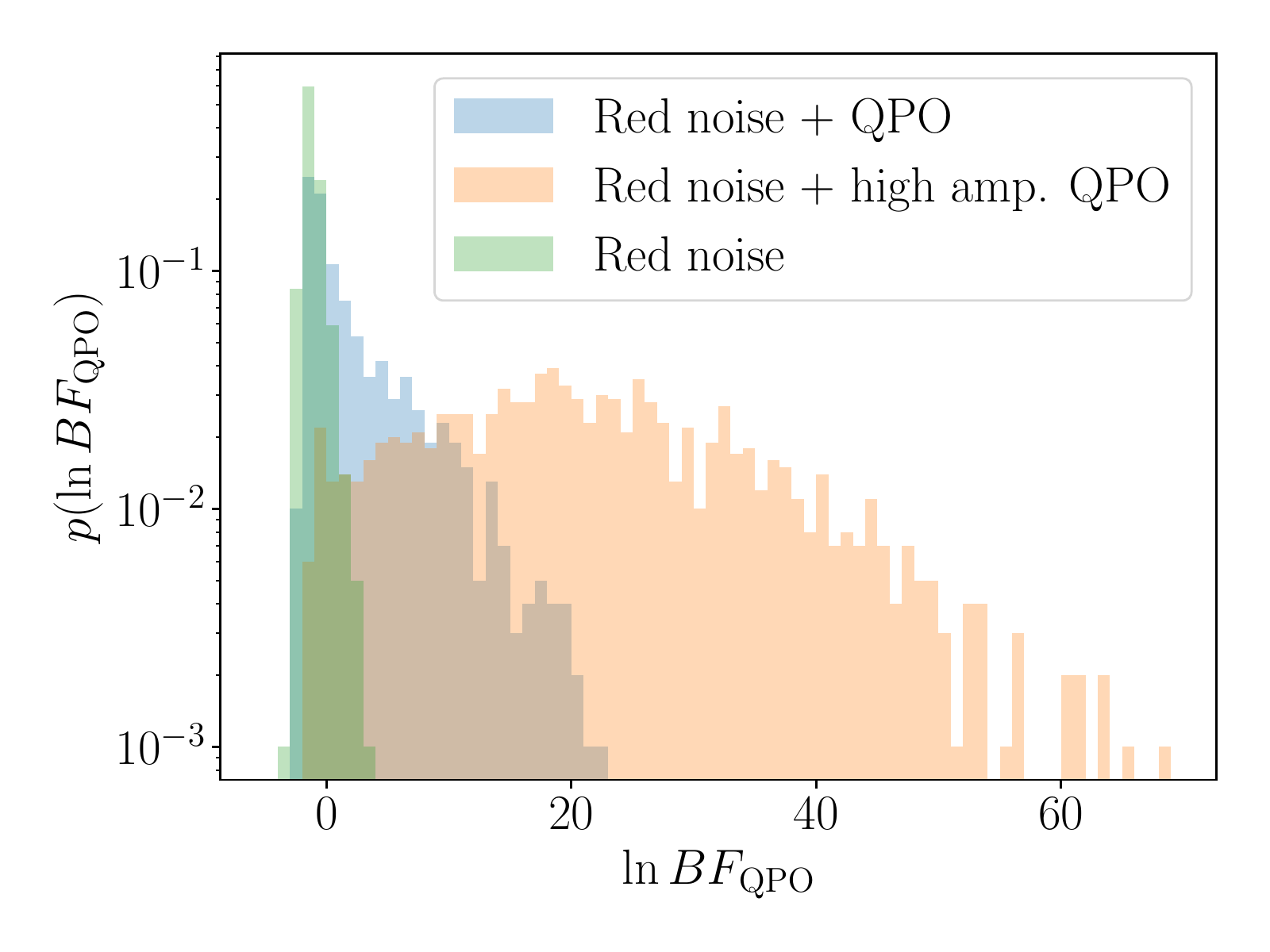}
    \caption{Normalised histogram of the $\ln \BFQPO$ values obtained from data sets produced with identical $\krnqpo$ (blue), $\krnqpo$ with a higher amplitude QPO (orange) and $\krn$ (green). 
    $52.9\%$ and $97.2\%$ of the lower and higher amplitude QPOs yield $\ln \BFQPO > 0$ (the logarithmic scale is somewhat deceiving).
    This shows that there is substantial spread in terms of the possible Bayes factors.}
    \label{fig:identical_injections_comparison}
\end{figure}

\begin{figure*}
    \centering
    \includegraphics[width=\textwidth]{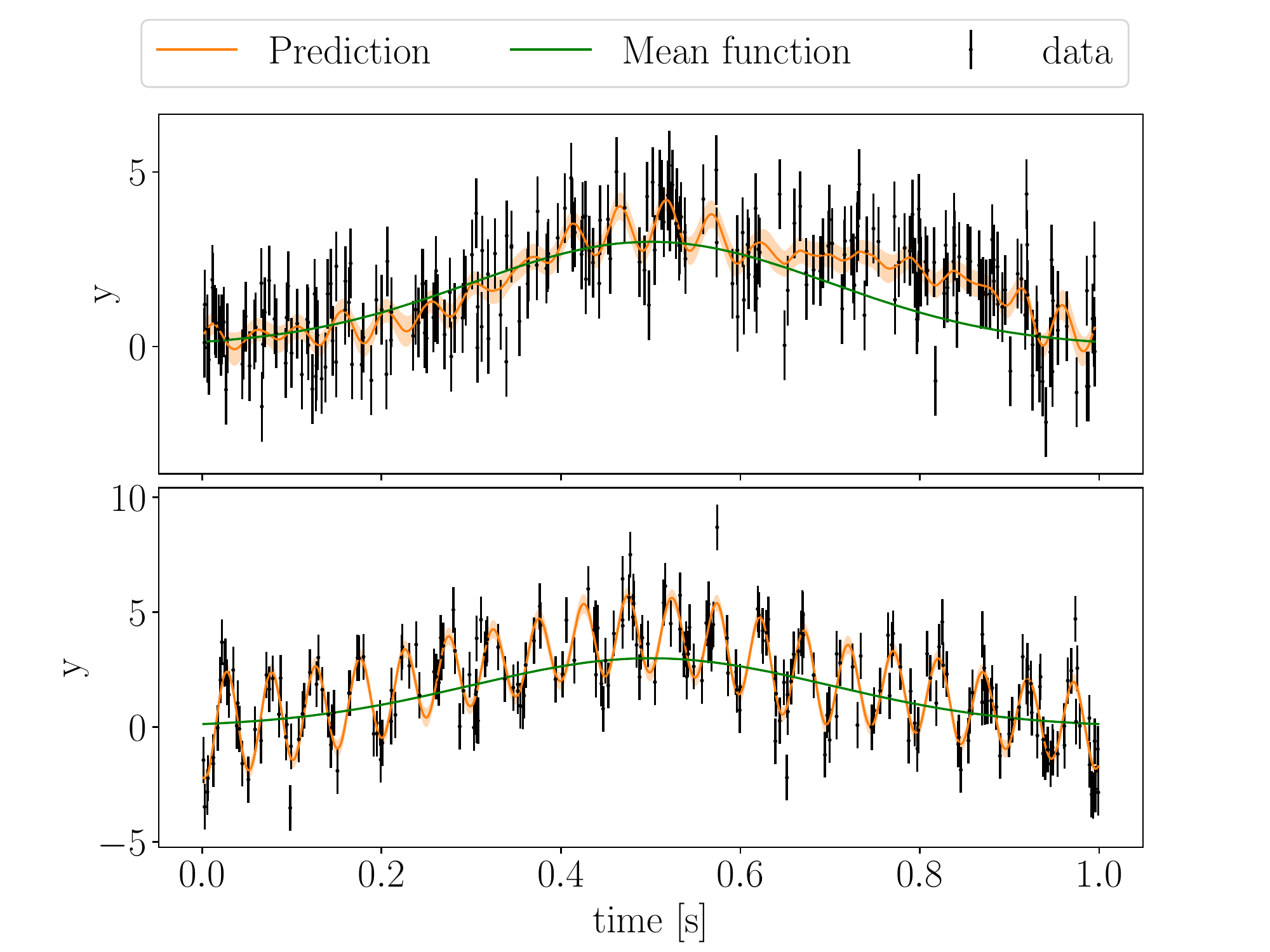}
    \caption{The least (top) and most (bottom) significant QPO in the higher QPO amplitude sets in the model selection study in Sec.~\ref{sec:model_selection}. 
    Despite having the same underlying parameters, the QPO is visibly much more pronounced in the bottom panel.}
    \label{fig:mss_extreme_cases}
\end{figure*}

In Fig.~\ref{fig:identical_injections_comparison} we show the result of performing this analysis for 1000 simulated light curves produced for each set of parameters.
The light curves containing a low amplitude QPO have a positive $\ln \BFQPO$ $56.7\%$ of the time, but some light curves yield $\ln \BFQPO > 20$, indicating very high significance.
High amplitude QPOs on the other yield a positive $\ln \BFQPO$ $ 97.2\%$ of the time, which indicates that we can almost always correctly identify QPOs with a sufficient amplitude.
Even for high amplitude QPOs the Bayesian analysis sometimes favours $\krn$ due to its smaller prior volume.
Figure~\ref{fig:mss_extreme_cases} demonstrates why there is such a high variance in the $\ln \BFQPO$ values we obtain.
Despite being drawn from the same multivariate Gaussian distribution, QPOs can have vastly different amplitudes in the actual light curves and even high amplitude QPOs can be missed.

On the other hand, $94\%$ of the data sets containing only red noise yield a $\ln \BFQPO < 0$, though the distribution is much narrower.
Only one simulated red noise data set achieved a $\ln \BFQPO > 2$.
This narrower distribution is likely because $\krn$ is a limiting case of $\krnqpo$.
A data set created using $\krn$ thus always fits well with $\krnqpo$, and the $\krn$ kernel model achieves a preference due to its smaller prior volume.
On the other hand, if there is significant oscillatory behaviour, $\krn$ cannot provide a good fit.

Testing the impact of modifying $\cqpo$ and $\fqpo$ on the expected significance in the form of a $\ln \BFQPO$ is expensive to perform empirically, so we will focus on a qualitative discussion.
Generally, a smaller $\cqpo$ corresponds to a higher quality QPO which should be more easily identifiable and thus yield a higher $\ln \BFQPO$. 
Higher frequency QPOs should also be more easily identifiable.
At higher frequencies, the QPO has more total periods within a fixed length light curve.
Additionally, if $\cqpo$ is fixed, the QPO will decay by less within a single period and thus have a higher oscillation quality.
This also highlights that identifying low-frequency QPOs is difficult because there are too few recorded oscillation periods within the data.

Overall, the results indicate that if we find a $\ln \BFQPO > 2$, it is very unlikely that we have seen a false positive.
We are also far more likely to miss real signals with weak QPOs than find false positive QPOs.
This tilt is preferable if we apply the methods on data where a false positive would be particularly detrimental for our understanding of astrophysics.
For example, if we consider astrophysical sources for which we do not yet know if they may contain QPOs or not.

\begin{table}[ht]
    \centering
    \begin{tabular}{c|c|c}
        Parameter   & $\krn$ values & $\krnqpo$ values \\ \hline\hline
        $\arn$      & $\exp(1)$     & $\exp(1)$ \\ \hline
        $\crn$      & $\exp(1)\, \si{\per\second}$     & $\exp(1)\, \si{\per\second}$ \\ \hline
        $\aqpo$     & -             & $\exp(-2)$ or $\exp(-0.4)$\\ \hline
        $\cqpo$     & -             & $\exp(1)\, \si{\per\second}$ \\ \hline
        $\fqpo$     & -             & \SI{20}{\hertz} \\ \hline
        $A$     & 3             & 3 \\ \hline
        $t_0$     & \SI{0.5}{\second}             & \SI{0.5}{\second} \\ \hline
        $\sigma$     & \SI{0.2}{\second}             & \SI{0.2}{\second} \\ \hline
        
    \end{tabular}
    \caption{Simulation parameters for model selection study. $\aqpo$ is listed for both the low and high amplitude simulations.}
    \label{tab:mss_injection_priors}
\end{table}

\subsection{Non-stationary Time Series}\label{sec:non_stat_time_series}

\citet{Huebner2021} has shown that inappropriately applying periodogram-based methods on astrophysical time series can vastly overestimate their significance. 
Specifically, periodograms only yield statistically independent frequency bins if the time series is stationary and homoscedastic.
Bias occurs because bins close to the QPO frequency are not pairwise statistically independent.
Thus, any analysis that combines the power from multiple bins that make up the QPO is bound to overestimate the QPO's significance.

Since the Whittle likelihood directly derives from the assumption that the time series is Gaussian, it is intuitive that the non-stationarity bias should also exist in GP likelihoods.
Gaussian Processes account for heteroscedasticity and deterministic trends, but the stationary kernels we are using cannot account for non-stationary behaviour in the noise as described in \citet{Huebner2021} with periodograms.
We laid out in Sec.~\ref{sec:c6:towards_non_stat_models} how we can create a simple, non-stationary GP model using \celerite.

\citet{Huebner2021} provides an extensive set of simulations to demonstrate the non-stationarity bias in simulated light curve.
We can use one of their light curves and results to perform a comparative study on how we can address the non-stationarity bias with GPs.
Specifically, \citet{Huebner2021} created the simulated light curve in Sec. 4.4 by producing red noise and a \SI{1}{\hertz} QPO from \SI{-10}{\second} to \SI{10}{\second} using the \citet{Timmer1995} method.
Next, they applied a Hann window to ensure a smooth turn on from zero and added Gaussian white noise between \SI{-200}{\second} to \SI{200}{\second} to create a non-stationary time series.
\citet{Huebner2021} showed that carrying out periodogram-based analysis on segments longer than the window containing the signal yield far higher $\ln \BFQPO$ due to the non-stationarity bias.

We apply the non-stationary version of the GP models from Sec.~\ref{sec:c6:towards_non_stat_models} and the regular stationary version on simulated data from Sec. 4.4 in \citet{Huebner2021}.
Thus, we can quantify the bias that occurs due to the non-stationary change in noise at $\pm\SI{10}{\second}$, as a function of the length of the white noise segments added on either side of the segment with the signal, parametrized as an extension factor $x$.

\begin{figure}
        \centering
        \includegraphics[width=\columnwidth]{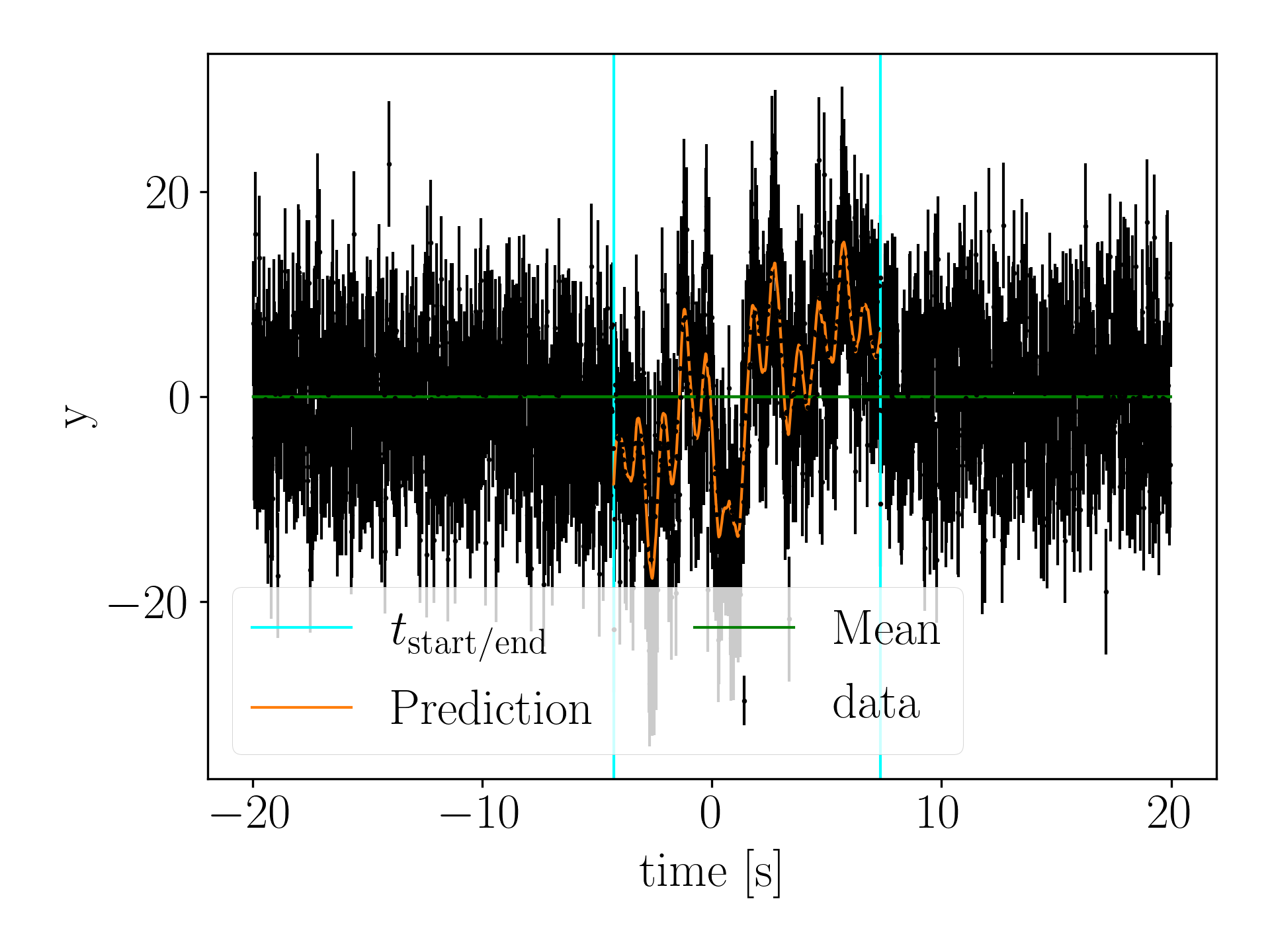} 
        \caption{Maximum likelihood fit of the data (black) from Sec. 4.4 in \citet{Huebner2021} with a non-stationary GP model.
        We use a constant zero mean model (green).
        The teal lines indicate the maximum likelihood $t_{\mathrm{start/end}}$ of the red noise and QPO GP.
        We underestimate the real duration of the GP which is \SI{20}{\second}.
        This is because \citet{Huebner2021} used a Hann window in the data creation which suppresses the amplitude closer to $\pm\SI{10}{\second}$.
        }\label{fig:non_stat_max_like_fit}
\end{figure}

\begin{figure}
    \centering
        \includegraphics[width=\columnwidth]{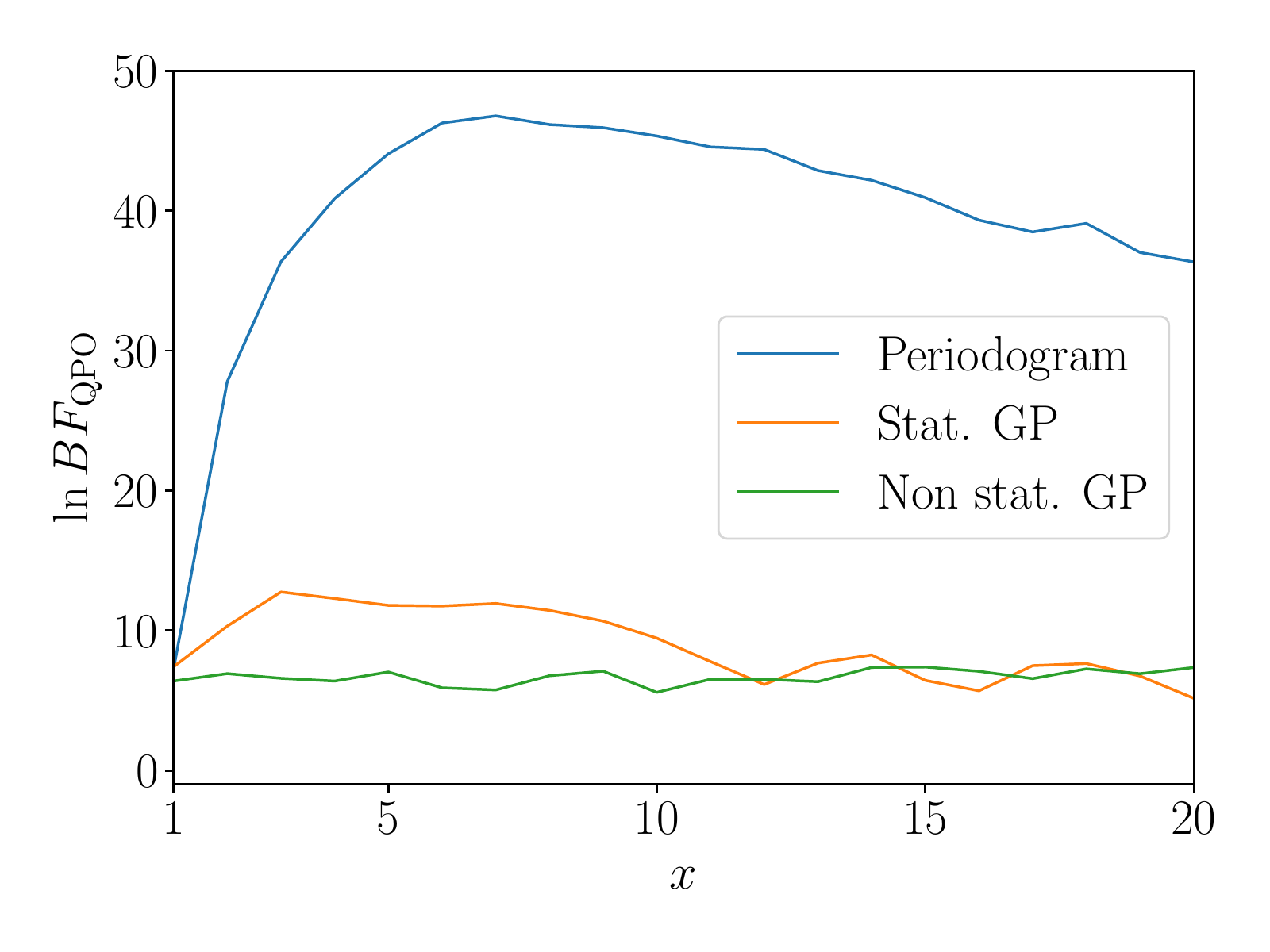} 
        \caption{$\ln \BFQPO$ for a segment extending from $-10x \, \mathrm{s}$ to $10x \, \mathrm{s}$ for a variable extension factor $x$. The signal is only contained in the central $20\,\mathrm{s}$, such that for $x=1$, the light curve contains only signal, for $x=2$, the light curve contains $10\,\mathrm{s}$ of white noise both before and after the segment with the signal, and so on. We compare three QPO detection methods: a standard periodogram-based method (blue), a stationary GP model (orange), and the non-stationary extension to the GP model (green). 
        While the periodogram and the stationary GP model are affected by the non-stationarity bias, the non-stationary GP model yields a near constant $\ln \BF$.}\label{fig:non_stat_comparison}
\end{figure}

\begin{figure}
        \centering
        \includegraphics[width=\columnwidth]{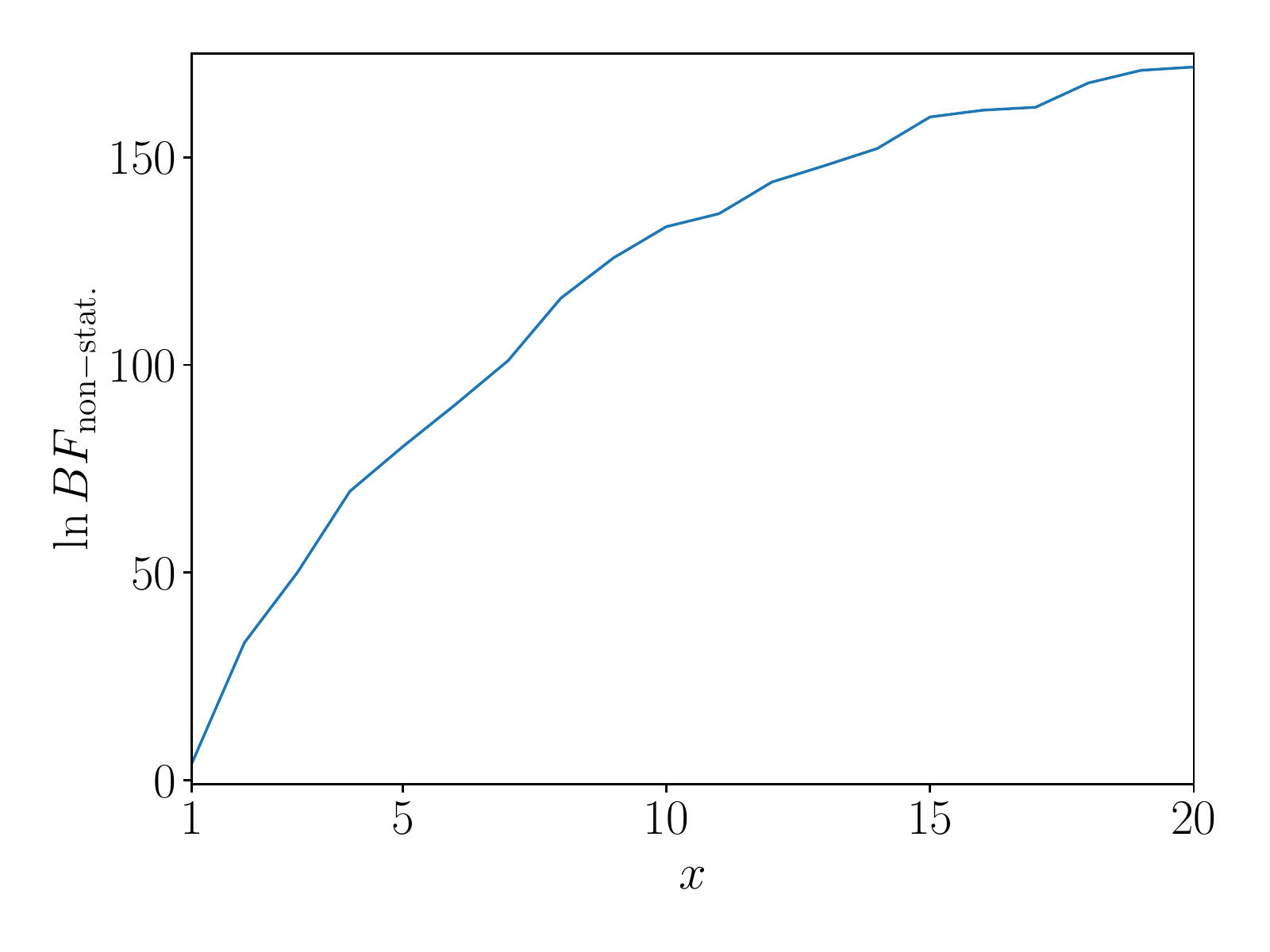}
        \caption{$\ln BF_{\mathrm{non-stat.}}$ of the non-stationary $\krnqpo$ GP model relative to the stationary GP for a segment between $-10x \, \mathrm{s}$ and $10x \, \mathrm{s}$. 
        The non-stationary model is preferred since it correctly models the change from red noise, white noise, and QPO to just white noise.}
        \label{fig:non_stat_vs_stat_gp}
\end{figure}

Figure ~\ref{fig:non_stat_max_like_fit} shows the maximum likelihood fit of the time series from \SI{-20}{\second} to \SI{20}{\second} using the non-stationary GP modification we introduce in Sec.~\ref{sec:c6:towards_non_stat_models}.
The $t_{\mathrm{start/end}}$ parameters delineate where the transition between GP and white noise occurs.
Because of this, the non-stationary GP model yields a near-constant $\ln \BFQPO$ regardless of how much the time series is extended by white noise, as we show in Fig.~\ref{fig:non_stat_comparison}.
This result for the non-stationary GP is desirable, since adding white noise on either end of a time series should not alter the information content about the QPO in the time series.
On the other hand, the $\ln \BFQPO$ values that we obtain for stationary GP models or the periodogram rises first due to non-stationarity bias and then falls off again as the QPO is covered by increasing white noise.
Empirically, we find that the bias is much stronger for periodograms.
This difference is likely because the kernel functions in the GP models effectively restrict the bias to a local time scale of $1/\cqpo$, whereas the periodogram uses a Fourier transform which calculates the power in each frequency bin from the entire time series.
Thus, far away from where the QPO occurs, no additional bias can accrue for the GP because we compare model fit of $\krn$ and $\krnqpo$ on white noise data.
We show in Fig.~\ref{fig:non_stat_vs_stat_gp} that the non-stationary $\krnqpo$ GP is indeed preferred over the stationary except for the segment between $\pm \SI{10}{\second}$.

\begin{figure}
    \centering
\includegraphics[width=\columnwidth]{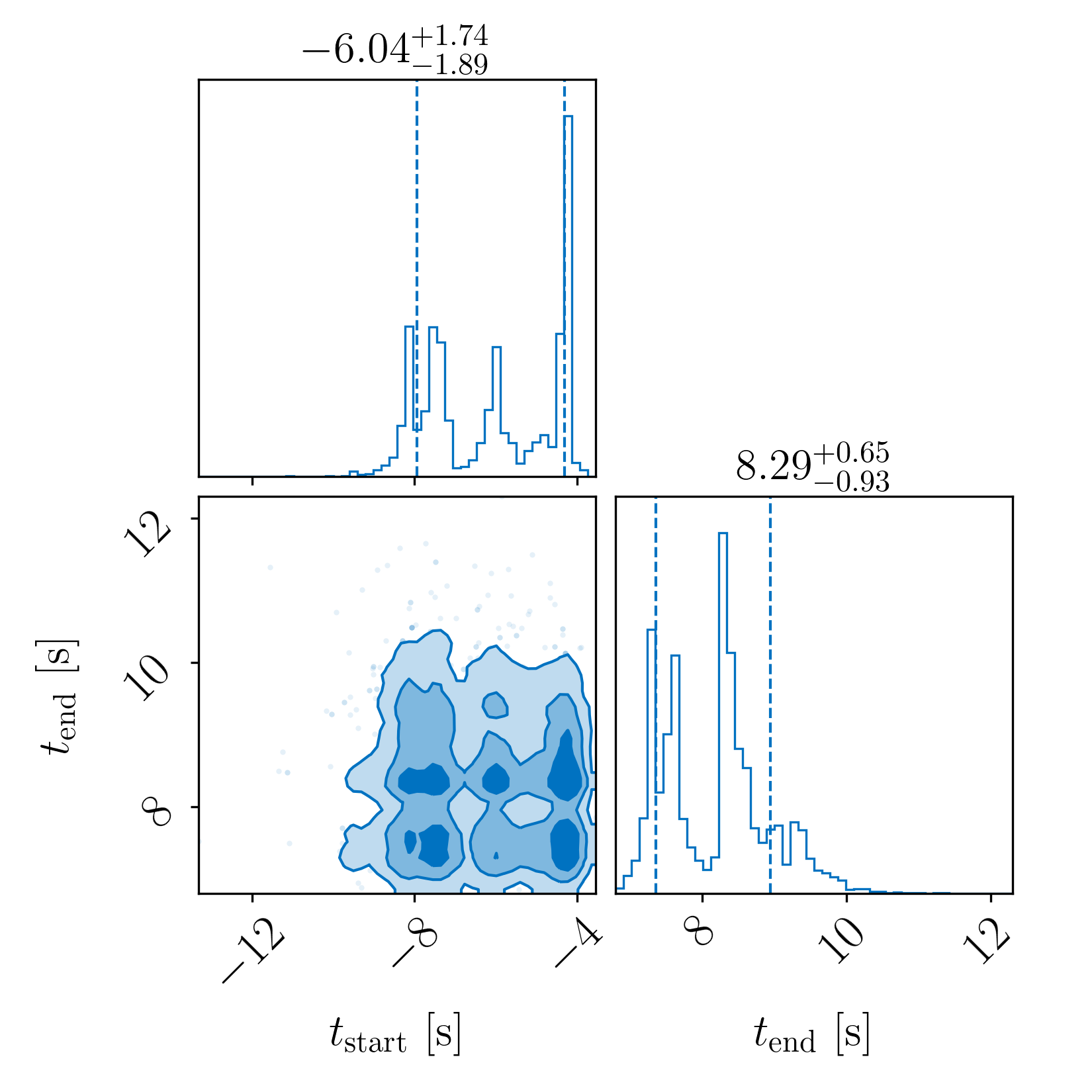}
        \caption{2D $t_{\mathrm{start}}$/$t_{\mathrm{end}}$ posterior for $x=2$ for the data from Sec. 4.4 in \citet{Huebner2021} which we show in Fig.~\ref{fig:non_stat_max_like_fit}.
        These parameters are hard to sample compared to the kernel and mean model parameters as they tend to be multimodal.}
        \label{fig:non_stat_start_end_time_corner}

\end{figure}

As we show in Fig.~\ref{fig:non_stat_start_end_time_corner} for the segment spanning between $\pm\SI{20}{\second}$, the two dimensional $t_{\mathrm{start/end}}$ posterior is well constrained. 
We note that these parameters have multimodal posterior distributions that are hard to sample.
The posterior we infer for $t_{\mathrm{start}}$ is also not consistent with the true start time of $-\SI{10}{\second}$ in \citet{Huebner2021}.
This inconsistency is likely because of how \citet{Huebner2021} created the data. Close to $\pm \SI{10}{\second}$ the Hann window suppresses red noise and QPO. 


\section{Solar Flares: Hares and Hounds}\label{sec:hares_hounds}

\begin{figure}
        \centering
        \includegraphics[width=\columnwidth]{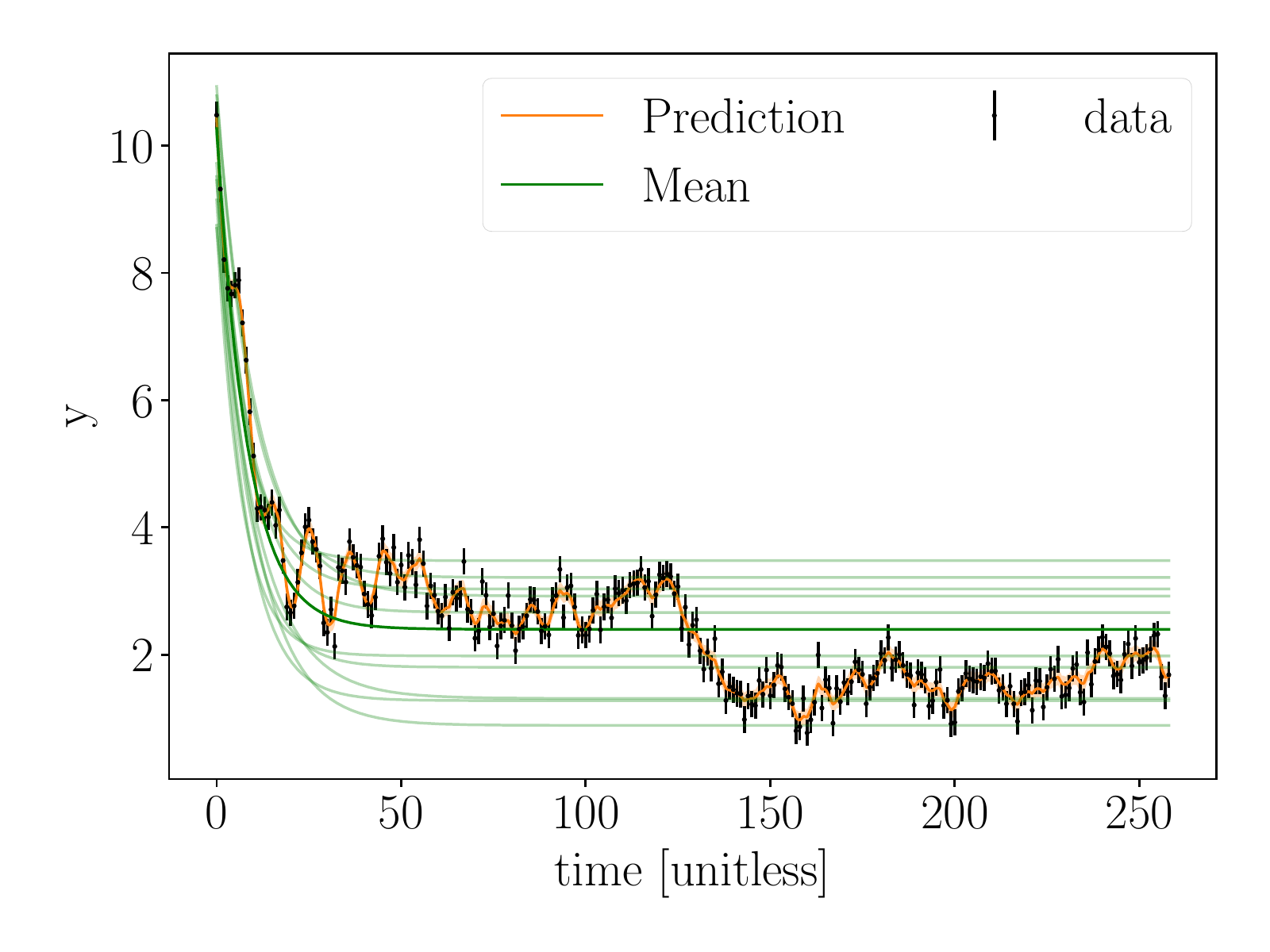}
        \caption{Maximum likelihood fit of the Hares and Hounds data set 404267 using a skewed exponential and the $\krnqpo$ kernel.
        The data contains a QPP of which a few cycles can be seen by eye for $t < 50$.
        We show the mean function from the maximum likelihood sample (dark green) and ten other samples from the posterior (light green). 
        The orange curve is the prediction based on the maximum likelihood sample and the 1-$\sigma$ confidence band.}
        \label{fig:hh_404267}
\end{figure}

Quasi-periodic pulsations  (QPPs)---as QPOs are called in the solar flare literature---have been a regularly reported feature of solar flares for decades (see~\citet{VanDoorsselaere2016, Zimovets2021} for recent reviews). 
They have also been observed, albeit less often, in stellar flare light curves. 
They can occur in a wide range of wavelengths and hence in both thermal and non-thermal flare emission, from radio waves (e.g.~\citet{Grechnev2003, Melnikov2005, Inglis2008, Kupriyanova2016}) to ultraviolet emission (e.g.~\citet{Brosius2016}), X-rays~\citep{Parks1969, Kane1983, Asai2001, Hayes2016, Hayes2019}, and even gamma-rays~\citep{Nakariakov2010}. 

On the Sun, the typical oscillation periods observed range from a few seconds up to several minutes. 
They can occur at any time during a flare, from the impulsive phase of energy release to the gradual decay phase. 
Quasi-periodic pulsations are particularly interesting observations in flares because they provide a potential diagnostic of fundamental flare energy release processes and the flaring plasma properties. 
The most established explanations for solar flare QPPs include periodic or bursty magnetic reconnection, and modulation of solar flare plasma via magnetohydrodynamic waves. 
A full discussion of potential emission mechanisms may be found in \citet{McLaughlin2018} and \citet{Zimovets2021}.

Historically, many different methods have been used to detect QPPs in solar flares. 
Recently, the robustness of different analysis methods has been studied by various authors (e.g.~\citet{Auchere2016, Dominique2018, Broomhall2019}). 
In particular, \citet{Broomhall2019} performed a comparative study between different QPP detection methods.
In this study, different methods were implemented and deployed by experts, who were blinded to which light curves did contain QPOS .
One of the methods used \celerite GP models in combination with a polynomial flare mean model~\citep{Davenport2014} and a detrending method.
The participants analysed 100 light curves; 60 contained a QPP.
The GP-based model in \citet{Broomhall2019} flagged 52 light curves for QPPs, only 29 of which actually contain a QPP, thus performed worse than picking events at random.
However, \citealt{Broomhall2019} used the stochastically-driven harmonic oscillator (SHO) kernel function as its sole model and implemented a heuristic detection criterion requiring that the oscillation period of the maximum likelihood estimate is between 3 and 200 units of time (data points are separated by 1 unit of time with 300 data points per flare). Without additionally including information about the quality factor $Q$ (i.e.~the width) of the QPO, it is unclear whether this condition alone could determine whether or not the SHO model is consistent with a QPO.

In the limit of low quality factors, the SHO becomes aperiodic, which should be used as a criterion to reject the QPP hypothesis.
Specifically, \citet{Foreman-Mackey2017} showed that the SHO has commonly used aperiodic forms in the limit of $Q \leq 1/2$ and $Q=1/\sqrt{2}$.
We conclude that the differences in performance between those in \citet{Broomhall2019} and our approach presented above are likely rooted in a combination of the different model assumption (e.g.~the detrending step) and the differences in how the significance is assessed.

We also note that the Hares and Hounds data diverge significantly from our assumptions about the data.
Firstly, \citet{Broomhall2019} created data with an unknown white noise component, so we add a white noise kernel term.
Secondly, the light curves are created with several different mean models.
Specifically, each simulation has flares based on Gaussian or exponential components, which are repeated once or twice in some cases, and the data has an offset and a linear trend.
Most importantly, while the QPP is intermittent and only appears on the tail side of the flare, red noise within these light curves persists for the entire segment.
The data is thus non-stationary in a way that we can not perfectly model using the modifications to \celerite we introduce in Sec.~\ref{sec:c6:towards_non_stat_models}.

\citet{Broomhall2019} uses an exponentially damped sinusoid as a QPP model.
Using an analytical function instead of a random process underlying the QPP also has implications for the results from the frequency-domain based methods in \citet{Broomhall2019}.
Specifically, frequency bins around the QPP frequency are not statistically independent.

Performing an analysis on all light curves can provide a point of comparison between our method and the different methods laid out in \citet{Broomhall2019}.
However, details such as mean models and noise properties were unknown to the participants in their study.

For our analysis, we restrict ourselves to a single skewed exponential flare mean model plus an offset and re-analyse the 100 light curves from the second Hares and Hounds round. 
We again compare the $\krn$ and $\krnqpo$ hypotheses using Bayesian inference, but we also add a white noise jitter term to the kernel to account for the unknown noise.
We also opt for removing the time series left of the maximum since the QPPs are explicitly only present in the decay phase.
Some study participants also made the same decision.
We display one of the fitted time series that contains a QPP in Fig.~\ref{fig:hh_404267}.

The data set contains 40 light curves with and 60 without a QPP. 
Additionally, there are some light curves with repeating flare shapes, and the flare shapes can either be Gaussian or exponential.

We find four light curves with $\ln \BFQPO \geq 2$, all of which contain a QPP. 
Given our findings in Sec.~\ref{sec:model_selection}, in which only one out of a thousand red noise signals caused a $\ln \BFQPO > 2$, this is a reasonable and conservative detection threshold.
The most significant QPP has $\ln \BFQPO = 9.7$.
There are three more light curves with $0 < \ln \BFQPO < 2$, none containing a QPP.
We thus find fewer QPPs than other models \citet{Broomhall2019} explored, specifically AFINO~\citep{Inglis2015, Inglis2016}, which found eight QPPs without a false detection.
However, the model setup here is a significant improvement over the implementation of the GP model in \citet{Broomhall2019}, suggesting that further development of our model taking into account the specific idiosyncracies of solar flare data will provide a powerful tool for QPP searches in these light curves.

\section{Real data}\label{sec:real_data}

\subsection{Gamma-ray Bursts}\label{sec:c6:grb}
There is ample speculation about the possibility for QPOs in long GRBs~\citep{Masada2007, Ziaeepour2011}, and there have been some recent claims about possible detections~\citep{Tarnopolski2021}.
One of the most tantalising events is GRB090709A, a long GRB for which \citet{Cenko2010} found a marginal $2\sigma$ \SI{8.06}{\second} period QPO in the prompt emission.
Independent analyses have since found similar significance levels between $2$ and $3.5\sigma$~\citep{Iwakiri2010, Ziaeepour2011, Dichiara2013, Guidorzi2016}.
GRB090709A had its strongest emission for about \SI{100}{\second} and a visible afterglow for several hundreds of seconds afterwards. 
\citet{Cenko2010} primarily relied on the \textit{Swift} light curve for their analysis, though they also considered the \textit{Suzaku} light curve, which did not increase the significance.
Further analyses of the \textit{Swift} and \textit{XMM-Newton} light curves by \citet{DeLuca2010}, who interpreted GRB090709A as a distant, standard, long GRB, also showed no periodicity above the $3\sigma$-level.
\textit{Konus} and \textit{SPI-ACS} also recorded light curves of GRB090709A, though it is not firmly established if the combined data of all instruments would yield higher significance~\citep{Iwakiri2010}.
\citet{Cenko2010} and \citet{DeLuca2010} detrend the light curve before calculating the periodogram to enhance the periodicity.

For our re-analysis of this event, we focus on the \textit{Swift} light curve, though joint inference of parameters across multiple observations is, in principle, possible.
We select $\SI{107}{\second}$ of the overall light curve as suggested by \citet{Cenko2010}.
We carry out inference with the models we define in Sec.~\ref{sec:mean_functions}, and allow a constant positive offset with all these models.
Furthermore, we analyse the data with 1-3 flare components to see if there is a substantial improvement when using a mean function with multiple components.
As with our framework described in~\ref{sec:kernel_func}, we carry out Bayesian model selection between $\krn$ and $\krnqpo$.

\begin{figure}
    \centering
    \includegraphics[width=\columnwidth]{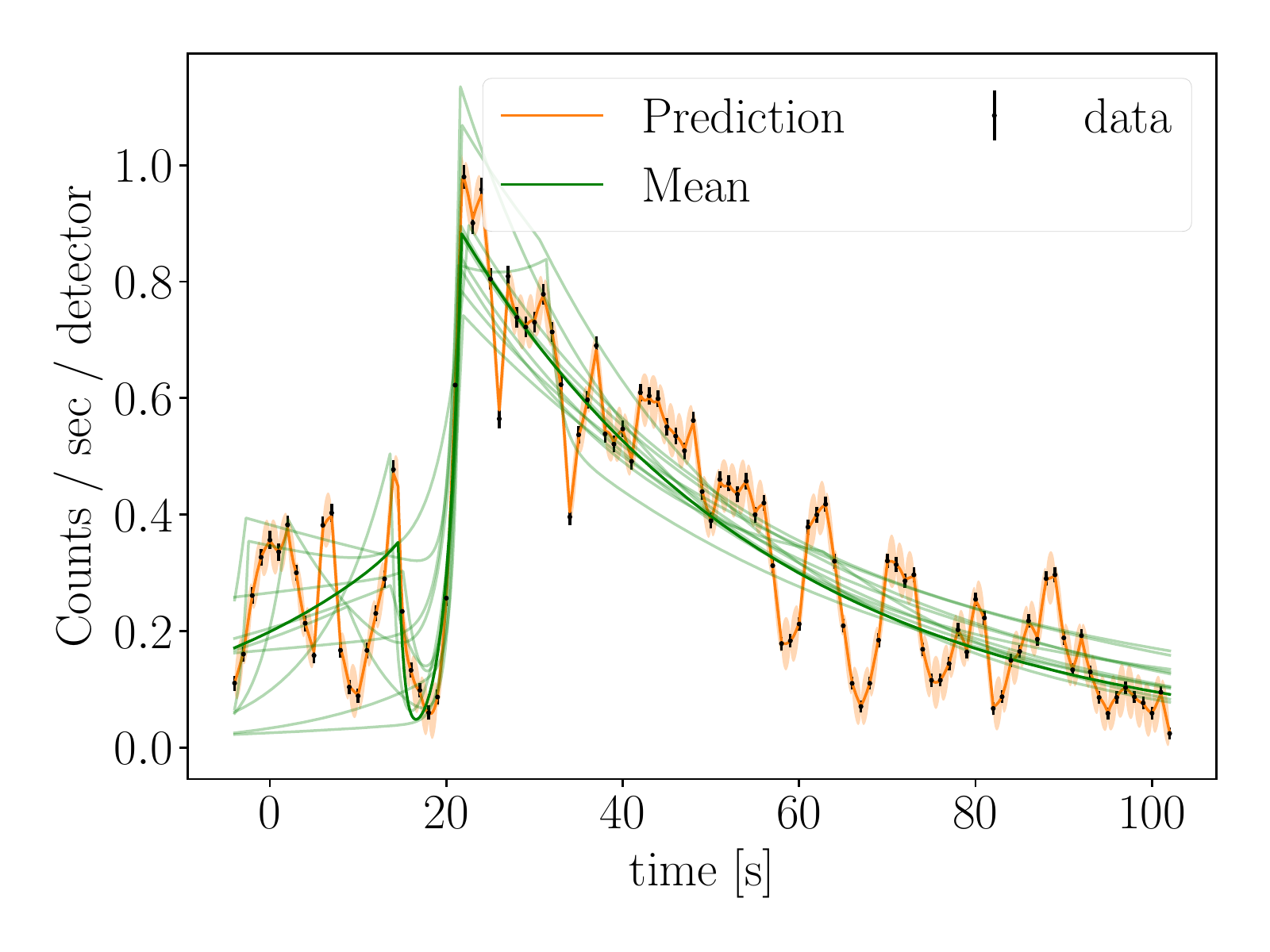}
    \caption{GRB090709A fit using two skewed exponentials and the $\krnqpo$ kernel.
    We show the mean function from the maximum likelihood sample (dark green) and ten other samples from the posterior (light green).
    The orange curve is the prediction based on the maximum likelihood sample and the 1-$\sigma$ confidence band.}
    \label{fig:GRB_max_like_fit}
\end{figure}

\begin{figure}
        \centering
        \includegraphics[width=\columnwidth]{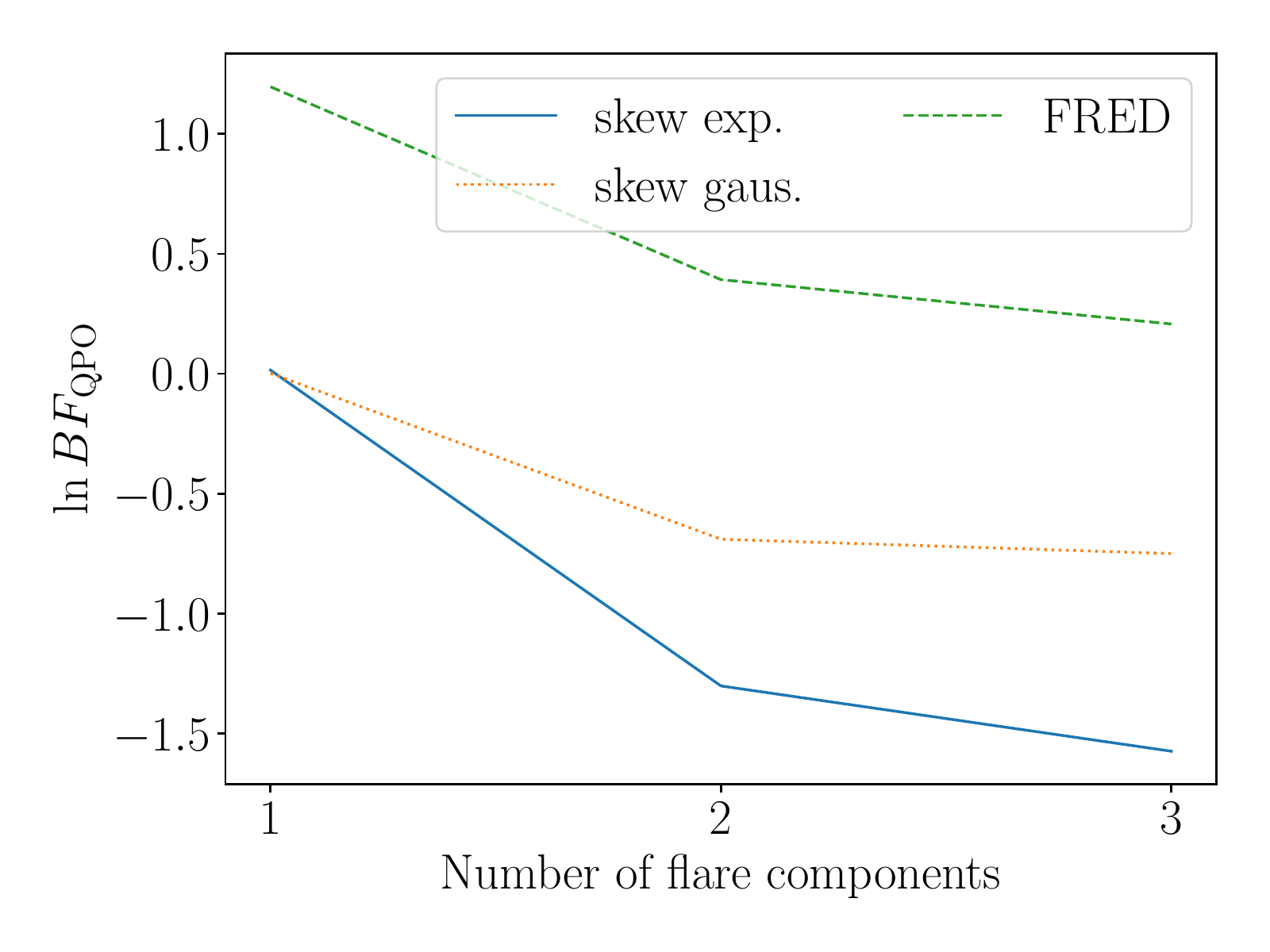}
        \includegraphics[width=\columnwidth]{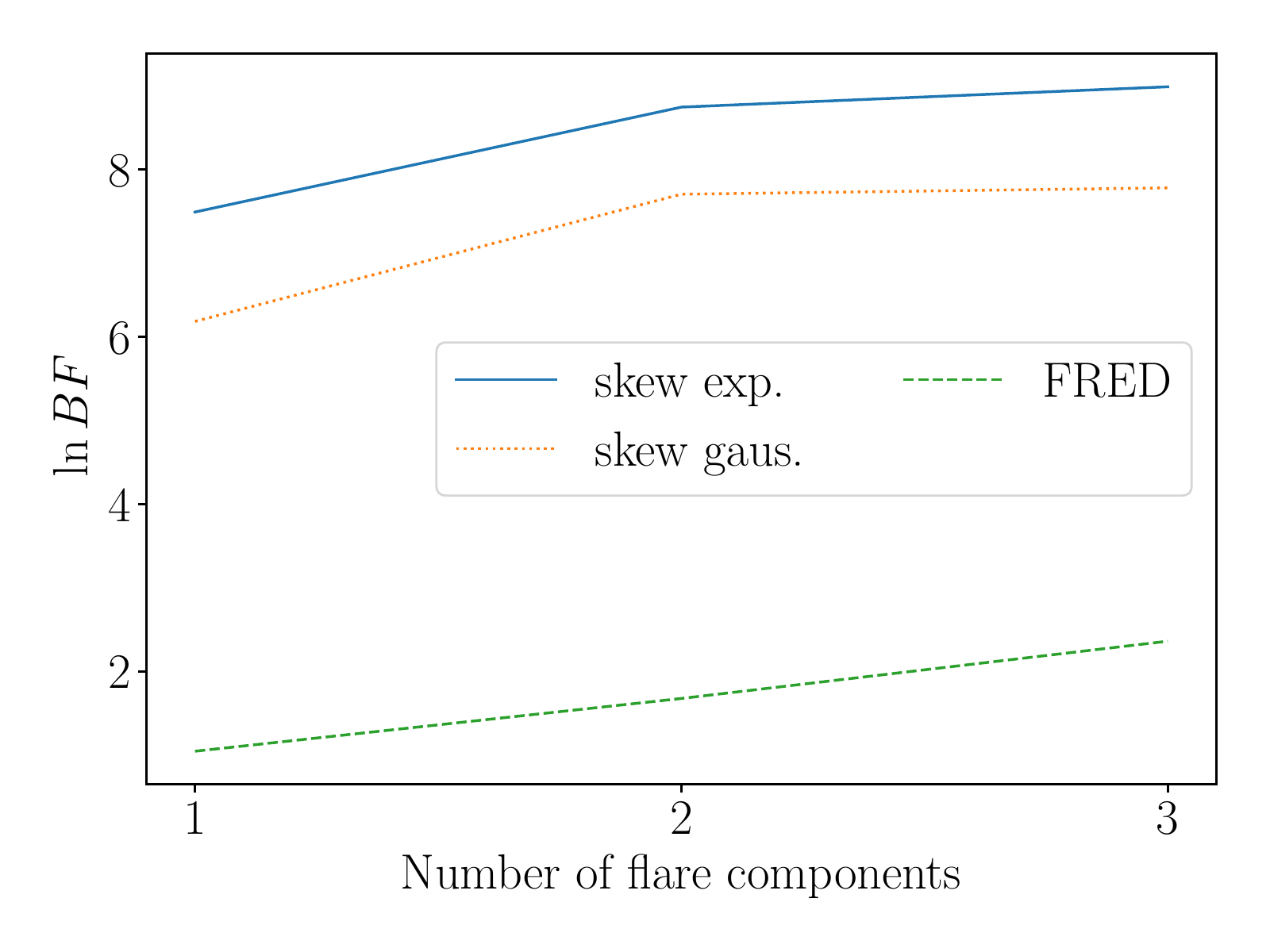}
        \caption{
        Results for GRB090709A.\\
        Top: $\ln \BFQPO$ for different mean models and different number of flare components. We find in all instances that neither $\krn$ nor $\krnqpo$ are strongly preferred for any specific configuration.\\
        Bottom: $\ln BF$ for different mean models using the $\krnqpo$ model relative to a constant zero mean model.}
        \label{fig:GRB_ln_BF}
\end{figure}

\begin{figure}
        \centering
        \includegraphics[width=\columnwidth]{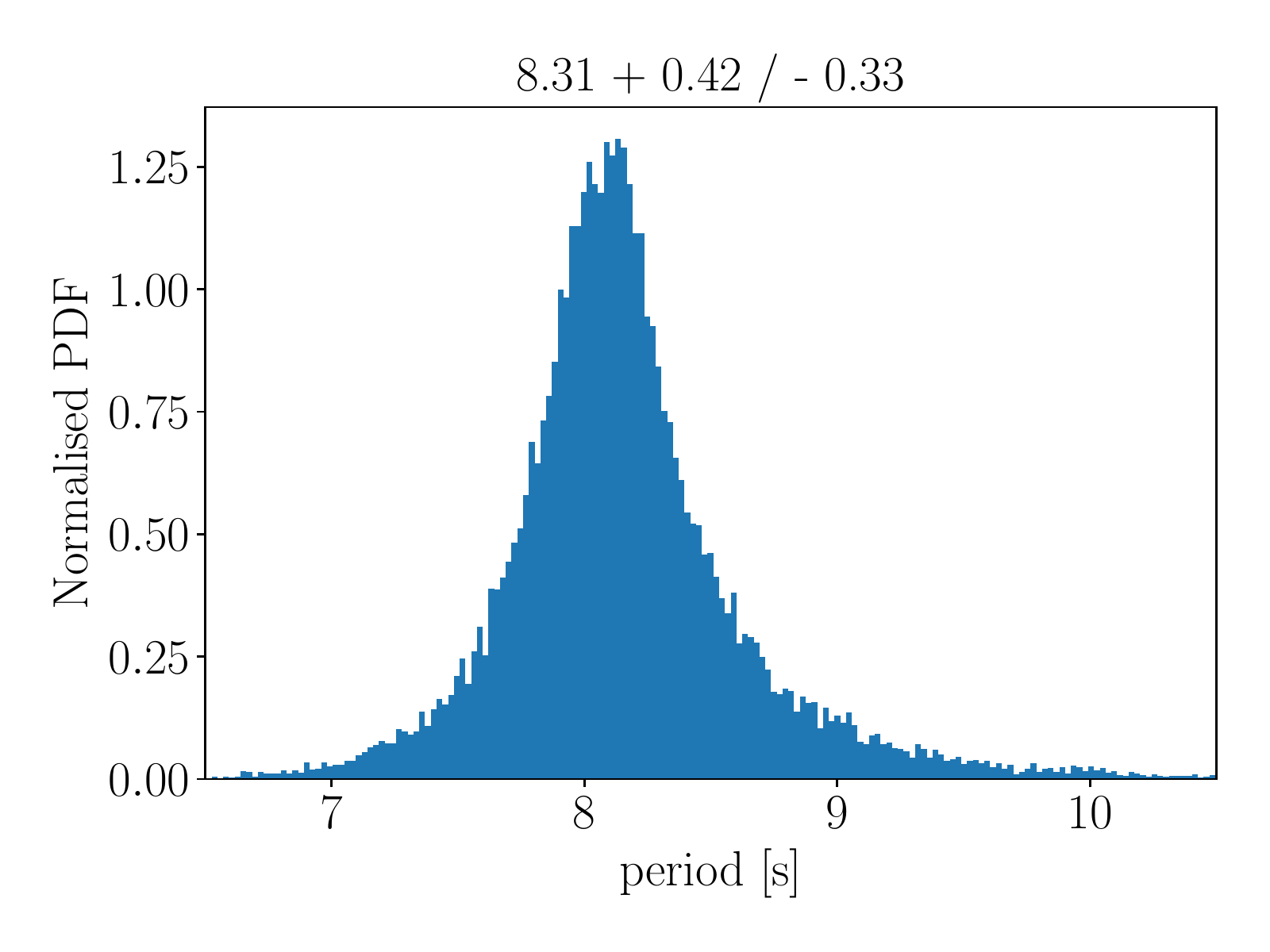}
        \caption{Period posterior for GRB090709A using two skewed exponentials. 
        The period is well constrained $P = 1/f = 8.31^{+0.42}_{-0.33}\si{\second}$ ($68\%$ CI) and consistent with the results from \citet{Cenko2010, DeLuca2010}, but has wide tails with low probability.}
        \label{fig:GRB_period_posterior}
\end{figure}

We perform Bayesian inference using \dynesty with 1000 live points and the random walk sampling method. 
Fig.~\ref{fig:GRB_max_like_fit} shows the data and a fit of the skewed exponential mean function with two flares as well as the prediction curve generated by \celerite using the maximum likelihood posterior sample.
In the top panel of Fig.~\ref{fig:GRB_ln_BF}, we display the obtained $\BFQPO$ for various mean models. 
We generally find that adding more components to our mean models does not meaningfully change the $\BFQPO$ and only somewhat increases the $\BFQPO$ relative to a constant zero mean model (bottom panel of Fig.~\ref{fig:GRB_ln_BF}).
This hints that already a single component model is a reasonable model for the overall trends in the data.
Furthermore, there is some difference between the evidences between the individual mean models.
The difference is relatively minor between the skewed exponential and Gaussian model, but the FRED model yields substantially lower evidences.
Overall, we find that the $\ln \BFQPO$ values falls between $-1.5$ and $1.1$.
We find that we obtain a well constrained $P = 1/f = 8.31^{+0.42}_{-0.33}\, \si{\second}$ ($68\%$ CI) posterior (see Fig.~\ref{fig:GRB_period_posterior}) that is consistent with the previously reported $P=\SI{8.1}{\second}$ period~\citep{Cenko2010, DeLuca2010}.
The low $\BFQPO$ indicate that we should not consider this to be a confident detection, though in the future, combining a significance criterion based on the Bayes factor with an \textit{effect size}, for example based on how well-constrained the posterior for the QPO frequency is, could improve our ability to detect QPOs in cases where the Bayes factor alone is inconclusive.


\subsection{Magnetar bursts}
Magnetars can show strong bursting behaviour, ranging from a series of low-energetic recurrent bursts to rare giant flares, of which we have observed few so far.
Some mechanisms may trigger a QPO in a magnetar burst, such as torsional Alfv\'en oscillations~\citep{Levin2007, Sotani2008}, but the observational evidence remains scant.
So far, there have only been definitive detections in the SGR1806-20, and SGR1900+14 giant flares~\citep{Strohmayer2005, Strohmayer2006, Israel2005, Watts2006} and in some recurring smaller bursts ~\citep{Huppenkothen2012, Huppenkothen2014, Huppenkothen2014b}. 

\begin{figure}
    \centering
    \includegraphics[width=\columnwidth]{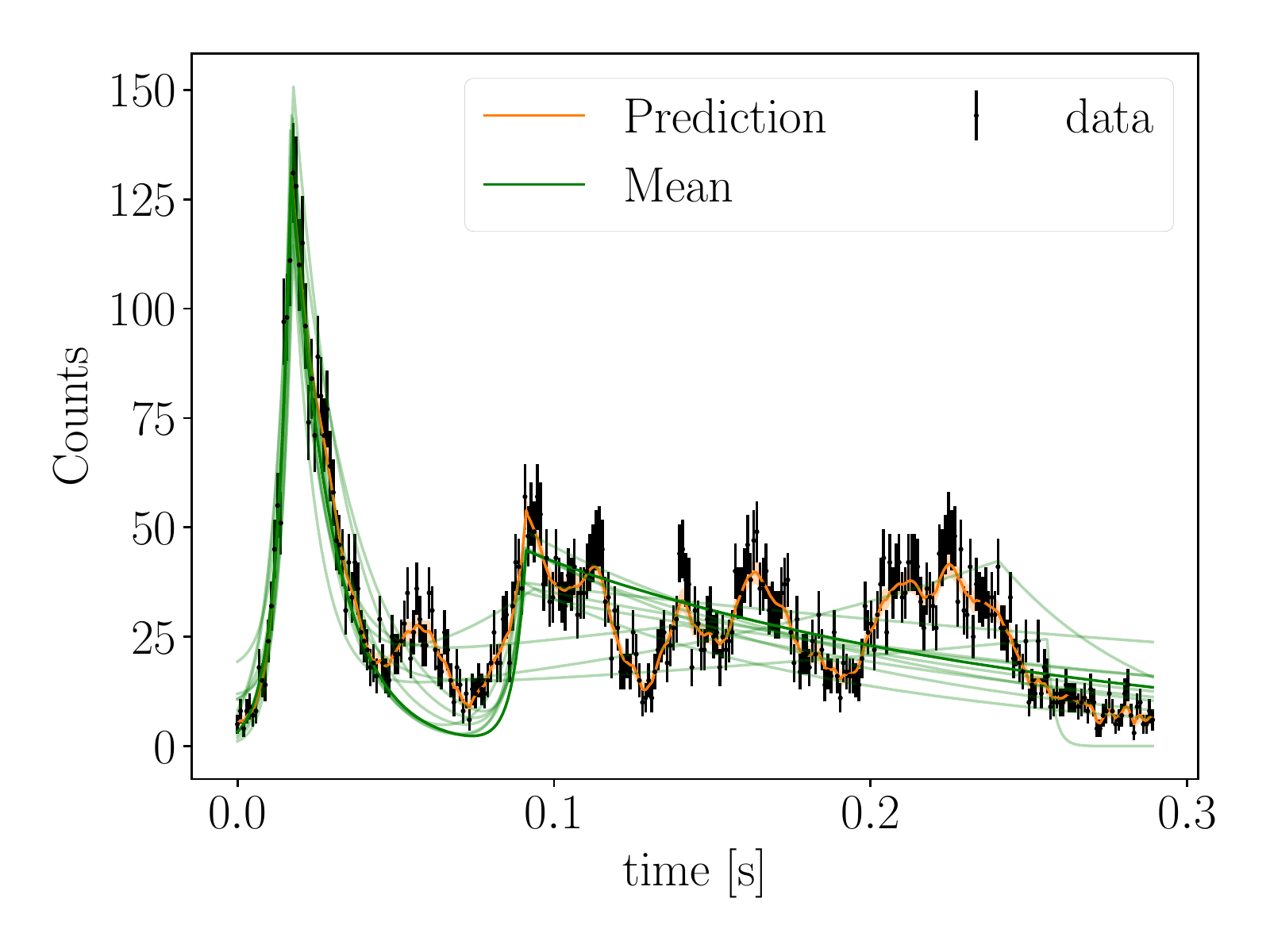}
    \caption{Magnetar burst 080823478 fit using two skewed exponentials and the $\krnqpo$ kernel.
    We show the mean function from the maximum likelihood sample (dark green) and ten other samples from the posterior (light green).
    The orange curve is the prediction based on the maximum likelihood sample and the 1-$\sigma$ confidence band.}
    \label{fig:magnetar_max_like_fit}
\end{figure}

\begin{figure}
        \centering
        \includegraphics[width=\columnwidth]{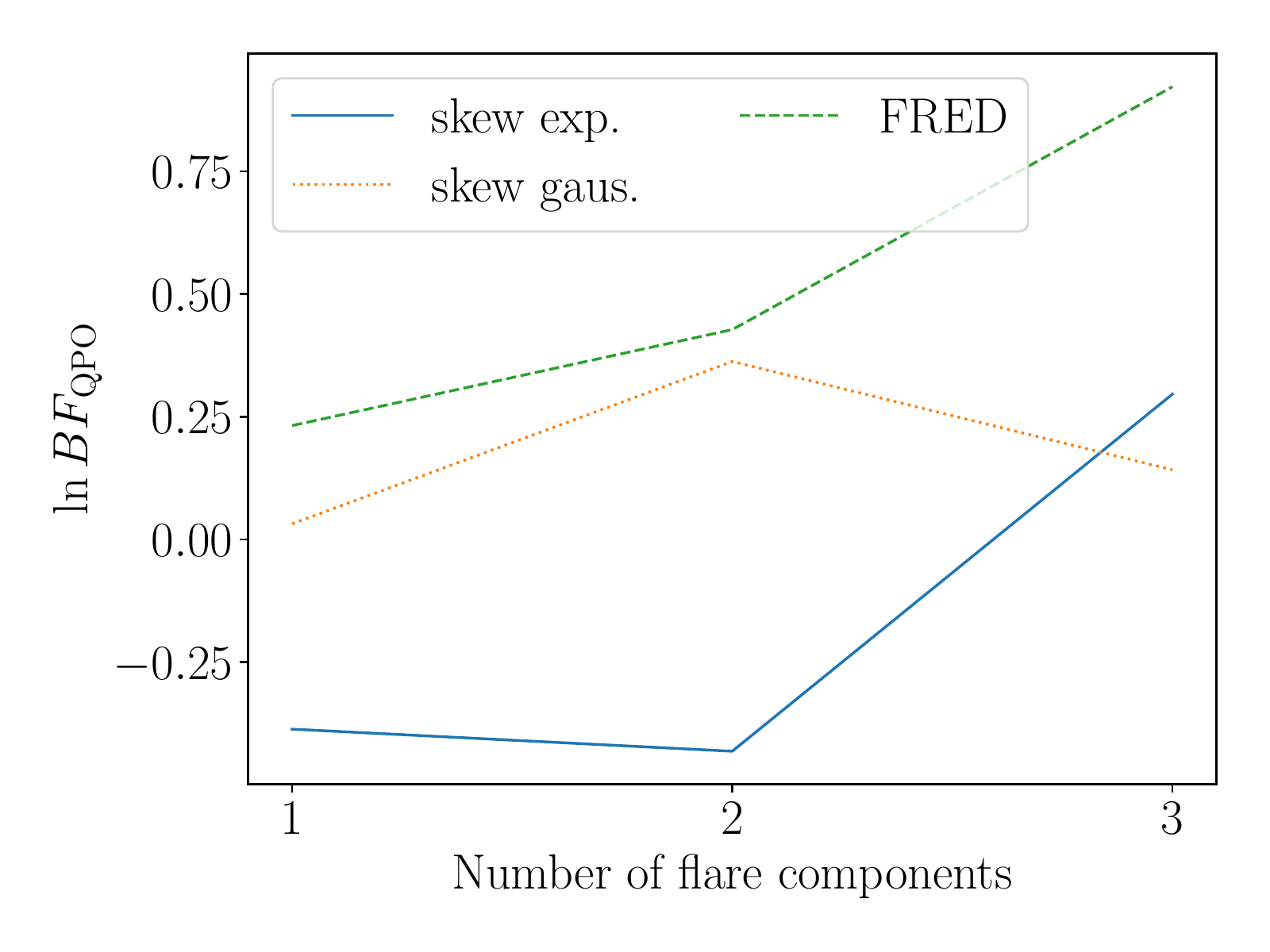}
        \includegraphics[width=\columnwidth]{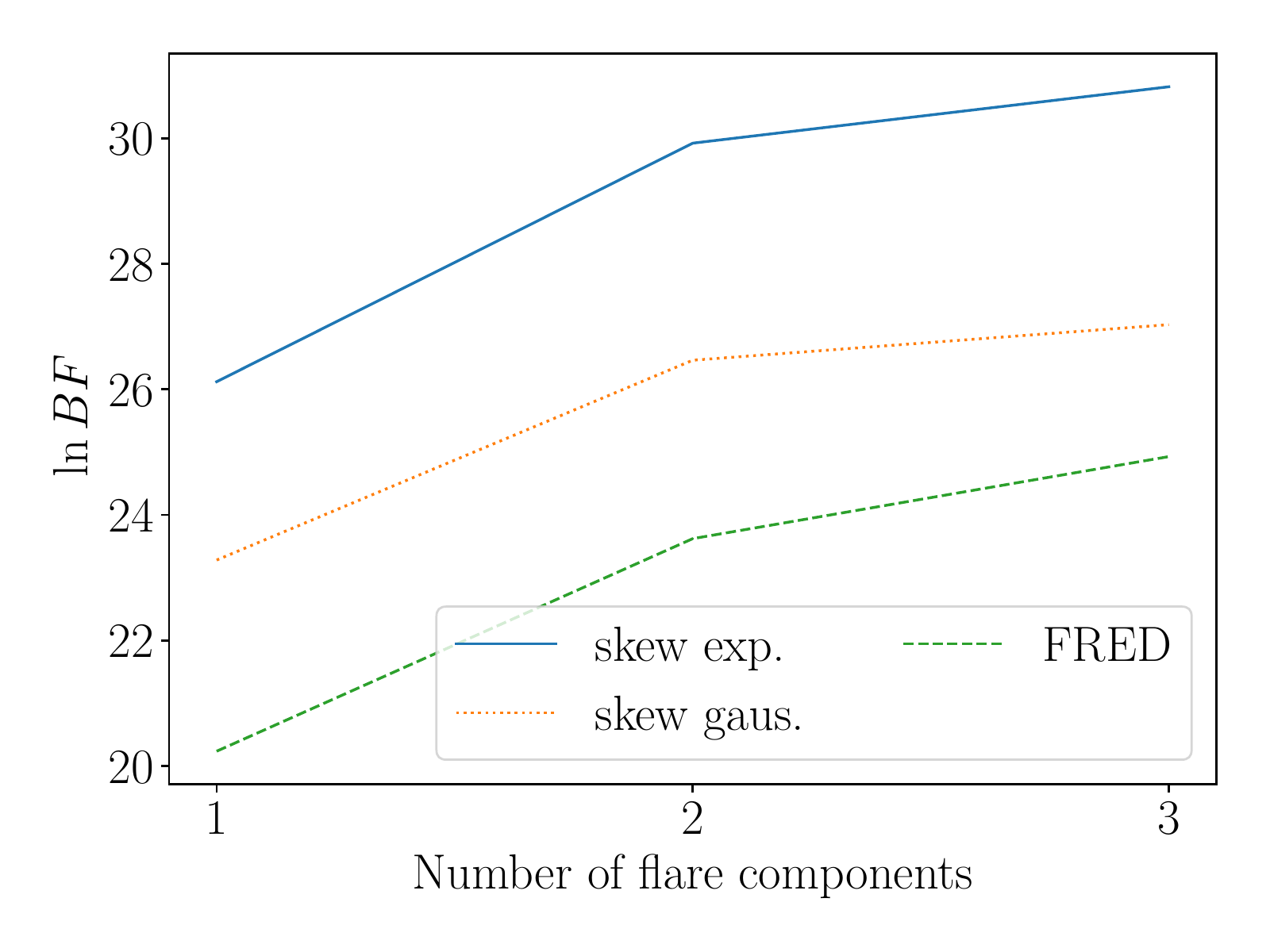}
        \caption{
        Results for the 080823478 burst.
        \\
        Top: $\ln \BFQPO$ for different mean models and different number of flare components. We find in all instances that neither $\krn$ nor $\krnqpo$ are strongly preferred for any specific configuration.\\
        Bottom: $\ln BF$ for different mean models using the $\krnqpo$ model relative to a constant zero mean model.}
        \label{fig:Magnetar_ln_BF}
\end{figure}

\begin{figure}
\centering
\includegraphics[width=\columnwidth]{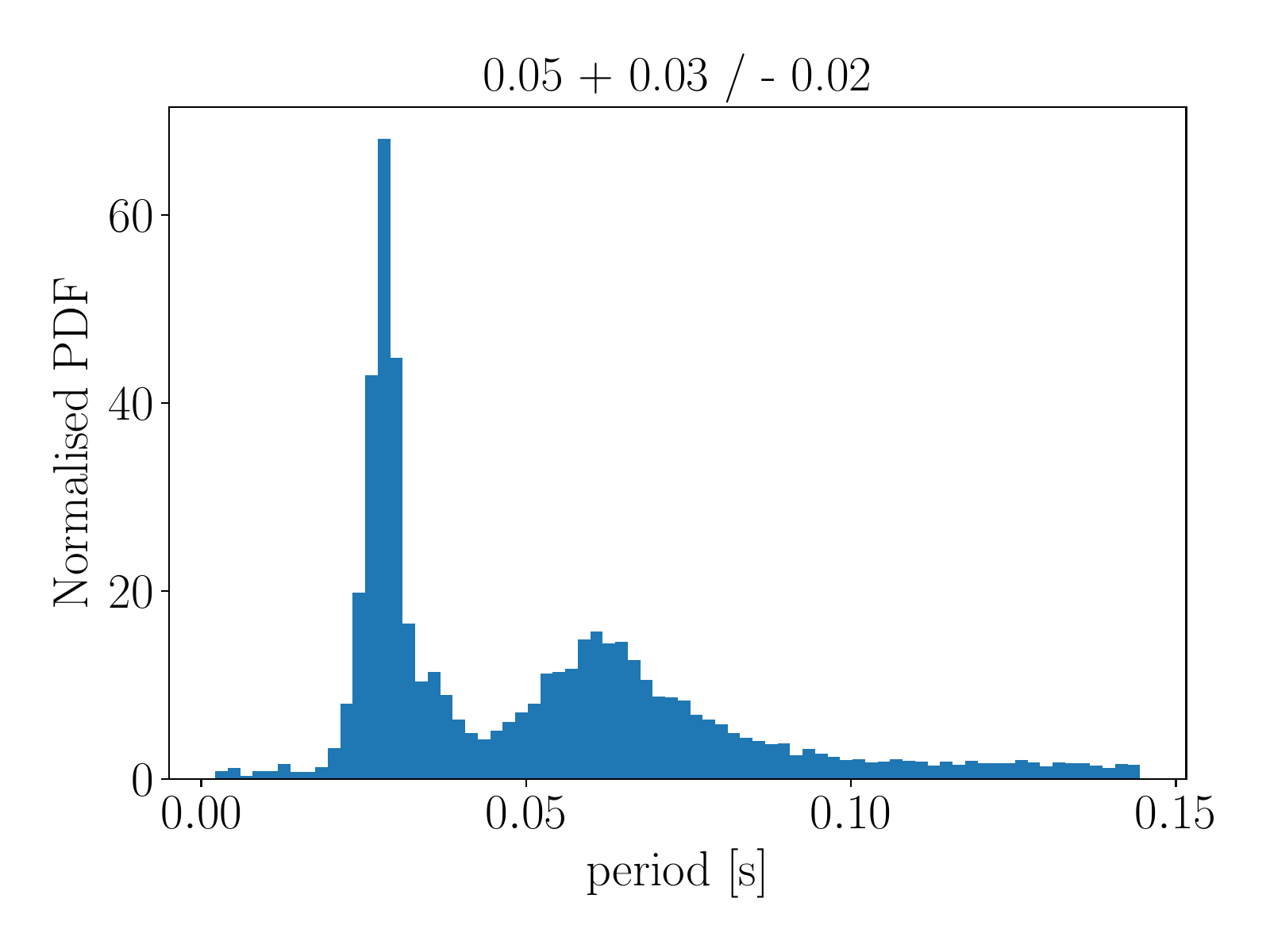}
    \caption{Period posterior for the 080823478 burst. The posterior is not constrained and has support across the entire prior space. This result is not representative of all runs we performed. Sometimes the period is more narrowly constrained, but $\krnqpo$ is still disfavoured.}
    \label{fig:magnetar_period_posterior}
\end{figure}

We demonstrate our method on a single magnetar burst 080823478 observed with the \textit{Fermi} Gamma-Ray Burst monitor from SGR0501+4516, a magnetar that was first discovered in 2008 by the \textit{Swift} satellite~\citep{Rea2009}. For details on the data processing, see \citet{Huppenkothen2012}.
As we can see in Fig.~\ref{fig:magnetar_max_like_fit}, the burst has a shape that, by eye, appears to be periodic, with three roughly equally spaced peaks after the main peak.
Following the same steps we have used for the GRB, we show in Fig.~\ref{fig:Magnetar_ln_BF} that the $\ln \BFQPO$ values vary between -0.4 and 0.9 depending on the mean model and there is thus no strong evidence for the presence of a QPO.
Moreover, the period posterior in Fig.~\ref{fig:magnetar_period_posterior} shows that we are also unable to consistently constrain $\fqpo$.
This finding is in agreement with the conclusions of \citet{Huppenkothen2012}, which used this burst as a template for studies of red noise in magnetar bursts and also found no periodicity.

Alternatively, we also test a constant mean model, meaning we model all variability as arising due to a stationary stochastic process.
In that case, $\krnqpo$ is strongly preferred over $\krn$ ($\ln \BFQPO = 6.8$), however, this mean model is strongly disfavoured compared to all the other mean models we have tested ($\ln BF = -21$ compared with the single skew exponential model).
These results highlight the benefit of using mean models as they help us disentangle whether variability arises from the overall shape of the burst or due to a stochastic process.

\subsection{Giant Magnetar Flare}
Multiple X-ray observatories, including the Rossi X-ray Timing Explorer (RXTE)~\citep{Hurley2005, Palmer2005} recorded the SGR1806-20 giant flare on December 27, 2004.
The flare, which lasted about \SI{380}{\second}, has been extensively studied for the presence of QPOs~\citep{Israel2005, Strohmayer2005, Strohmayer2006, Watts2006, Huppenkothen2014a, Miller2019}.
Notably, QPOs were associated with specific phases in the \SI{7.56}{\second} rotational period~\citep{Strohmayer2006}, which indicate that flare is associated with a specific region on the magnetar surface that is turning in and out of view.
\citet{Strohmayer2006} specifically identified the most significant QPOs at $\SI{18}{\hertz}$, $\SI{26}{\hertz}$, $\SI{29}{\hertz}$, $\SI{93}{\hertz}$, $\SI{150}{\hertz}$, $\SI{625}{\hertz}$, and $\SI{1837}{\hertz}$.
Initial attempts used averaged periodograms created by selecting the same time interval in subsequent rotational phases of the magnetar.
However, the non-stationary nature of the light curve within these time intervals may bias these results.

\citet{Miller2019} took a more systematic approach in analysing the giant flare by using a \SI{1}{\second} sliding window that they moved across the flare light curve in \SI{0.945}{\second} steps.
This way, they were able to locate when specific QPOs were occurring, though the time resolution is limited to $\approx \SI{1}{\second}$.

Given the findings of the non-stationarity bias in \citet{Huebner2021}, identifying when QPOs start and end in the flare light curve is an essential key to assessing their significance. 
We can use our non-stationary model from Sec.~\ref{sec:c6:towards_non_stat_models} to make quantitative statements for individual QPOs.

\begin{figure}
    \centering
    \includegraphics[width=\columnwidth]{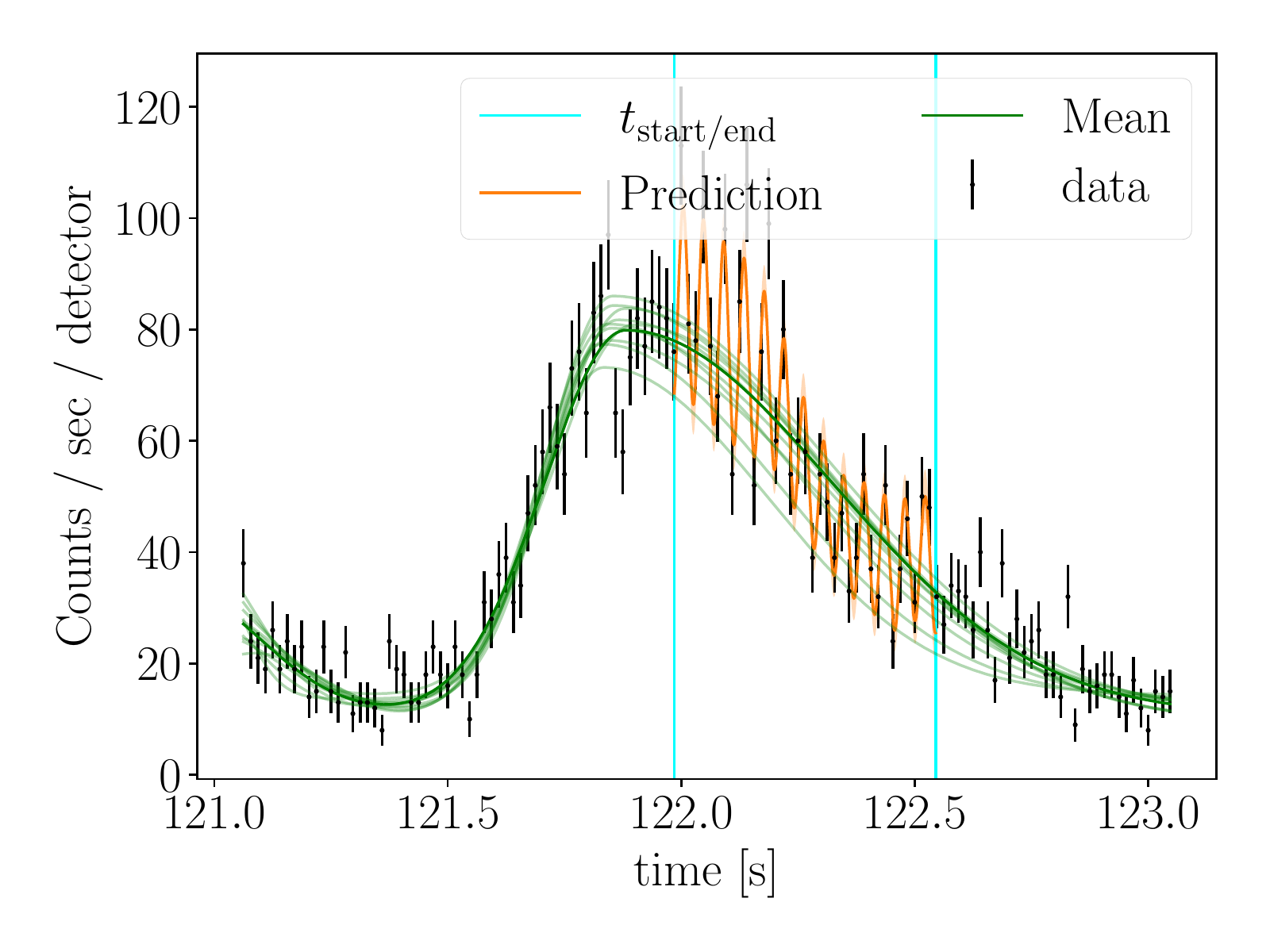}
    \caption{Giant flare maximum likelihood fit from our selected \SI{2}{\second} segment using two skewed Gaussians using the non-stationary $\krnqpo$ model.
    We show the mean function from the maximum likelihood sample (dark green) and ten other samples from the posterior (light green).
    The orange curve is the prediction based on the maximum likelihood sample and the 1-$\sigma$ confidence band.}
    \label{fig:giant_flare_max_like_fit}
\end{figure}

\begin{figure}
    \centering
    \includegraphics[width=\columnwidth]{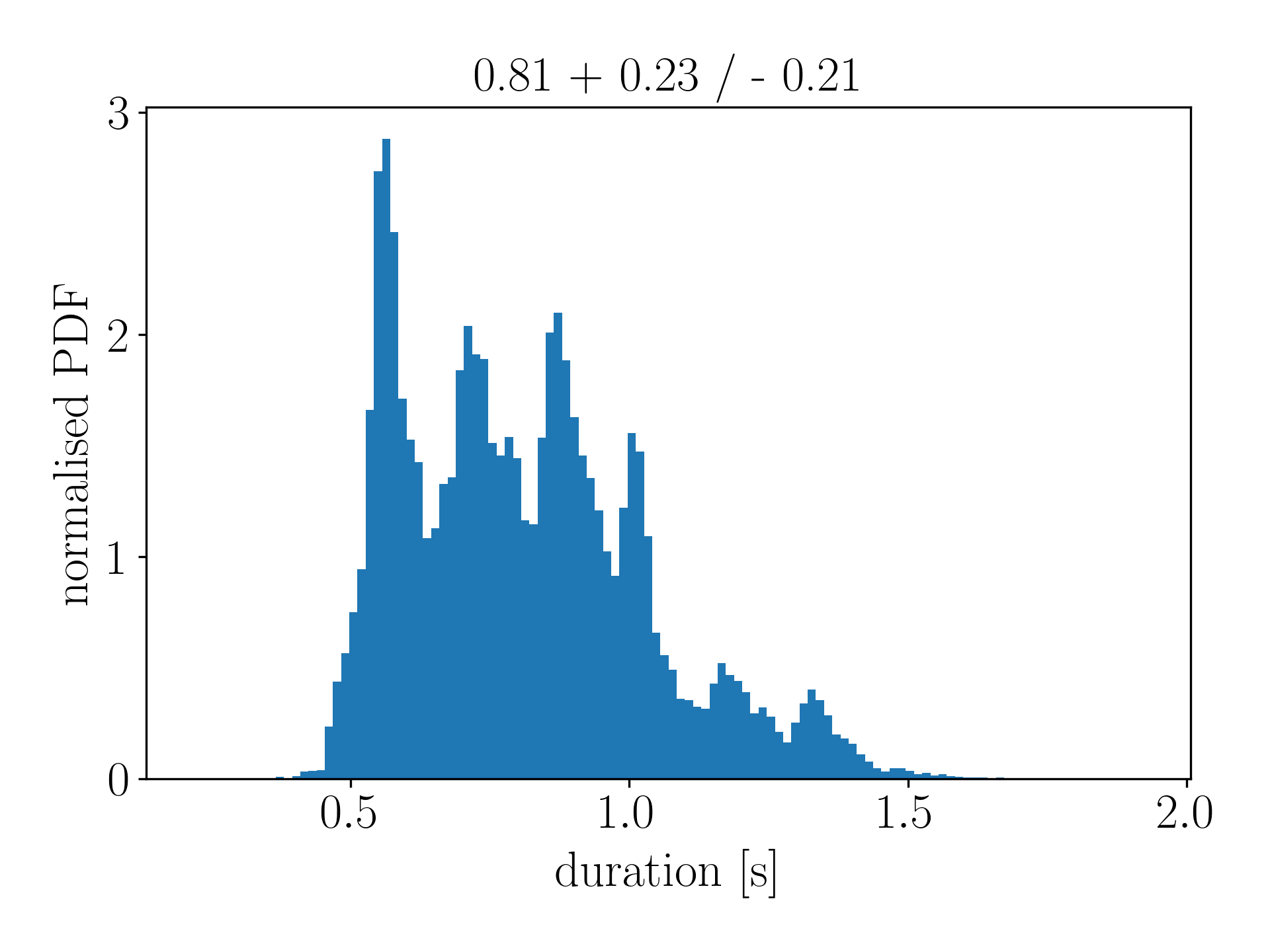}
    \caption{Posterior distribution of QPO duration $(t_{\mathrm{end}} - t_{\mathrm{start}})$ for the fit in Fig.~\ref{fig:giant_flare_max_like_fit}.} 
    \label{fig:giant_flare_duration_posterior}
\end{figure}

For this study, we focus on a specific \SI{23}{\hertz} QPO that was located in the RXTE light curve by \citet{Miller2019} to occur about \SI{122.060}{\second} after the beginning of the light curve at December 27, 2004, 21:30:31.378 UTC.
To see whether we can localise when the QPO occurs, we use the non-stationary model and select a \SI{2}{\second} segment starting at \SI{121.060}{\second}.
\citet{Miller2019} also reported a \SI{92}{\hertz} QPO to occur at the same segment.
Thus, we choose a relatively coarse binning of the time-tagged events of \SI{64}{\hertz} so that we can focus on the lower frequency QPO.
As we display in Fig.~\ref{fig:giant_flare_max_like_fit}, we can constrain the time when the QPO occurs within the light curve.
The inferred QPO mean frequency from the non-stationary model is $\fqpo = 22.90^{+0.45}_{-0.69}~\si{\hertz}$ ($68\%$ CI).
We find that the QPO lasts for $0.81^{+0.23}_{-0.21}\, \si{\second}$ ($68\%$ CI) and is primarily located on the top and tail side of the peak, though the posterior as shown in Fig.~\ref{fig:giant_flare_duration_posterior} is multi-modal with the maximum being closer to \SI{0.5}{\second}.
We obtain $\ln \BFQPO = 8.8$ using the non-stationary GP model, and $\ln \BFQPO = 6.8$ using the stationary model, and that the non-stationary $\krnqpo$ is preferred over the stationary one with $\ln BF = 5.1$.

Expanding on these results by analysing more segments may allow us to explore the temporal structure of the QPOs in more detail.
The specific segment we have considered contains one of the most significant QPOs in the entire flare.
More marginal QPOs are harder to find and characterise.
Eventually, hierarchical models may allow us to better understand the nature of QPOs in giant flares by taking a broader view of the entire \SI{400}{\second} of data.
Concretely, we may constrain where QPOs in the pulse period start and end and use information from multiple segments to find all frequency modes.

\section{Discussion and Outlook}\label{sec:discussion}
We introduce a new method to search for and analyse QPOs in the presence of red noise using a combination of GP modelling and Bayesian inference with evidence calculation.
We can model all aspects of an X-ray flare in the time domain using this method.
We use a set of phenomenological mean models to describe the overall flare shape and different GP models to model either red noise or a combination of red noise and QPOs.
Using studies on simulated data, we show that we can accurately estimate the true parameters, model simple non-stationary GPs, and avoid some of the biases that non-stationary behaviour causes when applying Fourier-based based methods.
We also show that we can distinguish QPOs from red noise, assuming the QPOs have a sufficient amplitude to be detected.
We demonstrate that one can easily apply this framework with few modifications on many different astrophysical X-ray transients.
Overall, the results from astrophysical data agree with the previously reported results in the literature.
We find that the application of GP methods in the analysis of X-ray time series is thus very promising, and specifically, we found that the method helps to constrain when QPOs occur within the light curve of SGR1806-20.,

While this paper provides an overview of how GPs can be applied in QPO searches in transient light curves, future work includes a number of improvements and extensions, in particular to make the method more powerful for specific sources and applications.
\celerite limits us to a narrow class of kernel functions, which we can extend by using a more general GP framework with fast solvers.
For example, the \textsc{HODLR} solver, of which the \textsc{george} package contains an implementation, operates in $\mathcal{O}(N\log^2 N)$  and thus can be run in acceptable time for the astrophysical data sets we have analysed~\citep{Ambikasaran2016}.
Specifically, this would allow us to implement kernels such as the Mat\'ern class of covariance functions or the squared exponential covariance function, which allows us to model a larger class of noise processes.
In Sec.~\ref{sec:limits}, we lay out assumptions underpinning and further limitations of GPs.
Specifically, we highlight the issues when analysing Poissonian data with low photon count rates, which has to be addressed using modified methods~\citep{Adams2009, Flaxman2015}. Extensions into non-stationary covariance functions would also be an interesting avenue for future work.

There have been recent claims of detections of QPOs in Fast Radio Bursts observed with the CHIME instrument~\citep{Andersen2021, PastorMarazuela2022}, which may be an interesting future application for the methods we have laid out in this paper.
\citet{PastorMarazuela2022} specifically considers FRB 20201020A for which they find a QPO at the $2.5\sigma$-level.
Fast radio burst studies currently use a mix of periodogram based methods and more bespoke methods for which a systematic application of GPs would provide a valuable cross-check.

\facilities{Fermi, Swift, RXTE}

\software{celerite \citep{Foreman-Mackey2017}, bilby \citep{Ashton2019,Romero-Shaw2020},dynesty \citep{speagle2020}, corner \citep{Foreman-Mackey2016}}

\begin{acknowledgments}
This work is supported through ARC Centre of Excellence CE170100004.
P.D.L. is supported through the Australian Research Council (ARC) Discovery Project DP220101610.
D.H. is supported by the Women In Science Excel (WISE) programme of the Netherlands Organisation for Scientific Research (NWO).
This work was performed on the OzSTAR national facility at Swinburne University of Technology. 
The OzSTAR program receives funding in part from the Astronomy National Collaborative Research Infrastructure Strategy (NCRIS) allocation provided by the Australian Government.
\end{acknowledgments}

\bibliography{QPOs}

\begin{thebibliography}{}
\expandafter\ifx\csname natexlab\endcsname\relax\def\natexlab#1{#1}\fi
\providecommand{\url}[1]{\href{#1}{#1}}
\providecommand{\dodoi}[1]{doi:~\href{http://doi.org/#1}{\nolinkurl{#1}}}
\providecommand{\doeprint}[1]{\href{http://ascl.net/#1}{\nolinkurl{http://ascl.net/#1}}}
\providecommand{\doarXiv}[1]{\href{https://arxiv.org/abs/#1}{\nolinkurl{https://arxiv.org/abs/#1}}}

\bibitem[{Adams {et~al.}(2009)Adams, Murray, \& MacKay}]{Adams2009}
Adams, R.~P., Murray, I., \& MacKay, D.~J. 2009, Proceedings of the 26th
  International Conference On Machine Learning, ICML 2009, 9

\bibitem[{Ambikasaran {et~al.}(2016)Ambikasaran, Foreman-Mackey, Greengard,
  Hogg, \& O'Neil}]{Ambikasaran2016}
Ambikasaran, S., Foreman-Mackey, D., Greengard, L., Hogg, D.~W., \& O'Neil, M.
  2016, IEEE Transactions on Pattern Analysis and Machine Intelligence, 38,
  252, \dodoi{10.1109/TPAMI.2015.2448083}

\bibitem[{Andersen {et~al.}(2021)Andersen, Bandura, Bhardwaj, Boyle, Brar,
  Breitman, Cassanelli, Chatterjee, Chawla, Cliche, Cubranic, Curtin, Deng,
  Dobbs, Dong, Fonseca, Gaensler, Giri, Good, Hill, Josephy, Kaczmarek, Kader,
  Kania, Kaspi, Leung, Li, Lin, Masui, Mckinven, Mena-Parra, Merryfield,
  Meyers, Michilli, Naidu, Newburgh, Ng, Ordog, Patel, Pearlman, Pen, Petroff,
  Pleunis, Rafiei-Ravandi, Rahman, Ransom, Renard, Sanghavi, Scholz, Shaw,
  Shin, Siegel, Singh, Smith, Stairs, Tan, Tendulkar, Vanderlinde, Wiebe, Wulf,
  \& Zwaniga}]{Andersen2021}
Andersen, B.~C., Bandura, K., Bhardwaj, M., {et~al.} 2021.
\newblock \doarXiv{2107.08463}

\bibitem[{{Angus} {et~al.}(2018){Angus}, {Morton}, {Aigrain}, {Foreman-Mackey},
  \& {Rajpaul}}]{angus2018}
{Angus}, R., {Morton}, T., {Aigrain}, S., {Foreman-Mackey}, D., \& {Rajpaul},
  V. 2018, \mnras, 474, 2094, \dodoi{10.1093/mnras/stx2109}

\bibitem[{Anscombe(1948)}]{Anscombe1948}
Anscombe, F.~J. 1948, Biometrika, 35, 246, \dodoi{10.2307/2332343}

\bibitem[{Asai {et~al.}(2001)Asai, Shimojo, Isobe, Morimoto, Yokoyama,
  Shibasaki, \& Nakajima}]{Asai2001}
Asai, A., Shimojo, M., Isobe, H., {et~al.} 2001, The Astrophysical Journal,
  562, L103, \dodoi{10.1086/338052}

\bibitem[{Ashton {et~al.}(2019{\natexlab{a}})Ashton, Lasky, Graber, \&
  Palfreyman}]{Ashton2019}
Ashton, G., Lasky, P.~D., Graber, V., \& Palfreyman, J. 2019{\natexlab{a}},
  Nature Astronomy, 3, 1143, \dodoi{10.1038/s41550-019-0844-6}

\bibitem[{Ashton {et~al.}(2019{\natexlab{b}})Ashton, H{\"{u}}bner, Lasky,
  Talbot, Ackley, Biscoveanu, Chu, Divakarla, Easter, Goncharov, Vivanco,
  Harms, Lower, Meadors, Melchor, Payne, Pitkin, Powell, Sarin, Smith, \&
  Thrane}]{Ashton2018}
Ashton, G., H{\"{u}}bner, M., Lasky, P.~D., {et~al.} 2019{\natexlab{b}}, The
  Astrophysical Journal Supplement Series, 241, 27,
  \dodoi{10.3847/1538-4365/ab06fc}

\bibitem[{Auch{\`{e}}re {et~al.}(2016)Auch{\`{e}}re, Froment, Bocchialini,
  Buchlin, \& Solomon}]{Auchere2016}
Auch{\`{e}}re, F., Froment, C., Bocchialini, K., Buchlin, E., \& Solomon, J.
  2016, The Astrophysical Journal, 825, 110,
  \dodoi{10.3847/0004-637X/825/2/110}

\bibitem[{Bar-Lev \& Enis(1990)}]{Bar-Lev1990}
Bar-Lev, S.~K., \& Enis, P. 1990, Statistics and Probability Letters, 10, 95,
  \dodoi{10.1016/0167-7152(90)90002-O}

\bibitem[{Bond {et~al.}(1999)Bond, Crittenden, Jaffe, \& Knox}]{Bond1999}
Bond, J., Crittenden, R., Jaffe, A., \& Knox, L. 1999, Computing in Science and
  Engineering, 1, 21, \dodoi{10.1109/5992.753044}

\bibitem[{Bond \& Efstathiou(1987)}]{Bond1987}
Bond, J.~R., \& Efstathiou, G. 1987, Monthly Notices of the Royal Astronomical
  Society, 226, 655, \dodoi{10.1093/mnras/226.3.655}

\bibitem[{Broomhall {et~al.}(2019)Broomhall, Davenport, Hayes, Inglis,
  Kolotkov, McLaughlin, Mehta, Nakariakov, Notsu, Pascoe, Pugh, \&
  Doorsselaere}]{Broomhall2019}
Broomhall, A.~M., Davenport, J.~R., Hayes, L.~A., {et~al.} 2019, arXiv,
  \dodoi{10.3847/1538-4365/ab40b3}

\bibitem[{Brosius {et~al.}(2016)Brosius, Daw, \& Inglis}]{Brosius2016}
Brosius, J.~W., Daw, A.~N., \& Inglis, A.~R. 2016, The Astrophysical Journal,
  830, 101, \dodoi{10.3847/0004-637X/830/2/101}

\bibitem[{Castro-Tirado {et~al.}(2021)Castro-Tirado, {\O}stgaard,
  G{\"{o}}ǧ{\"{u}}ş, S{\'{a}}nchez-Gil, Pascual-Granado, Reglero, Mezentsev,
  Gabler, Marisaldi, Neubert, Budtz-J{\o}rgensen, Lindanger, Sarria, Kuvvetli,
  Cerd{\'{a}}-Dur{\'{a}}n, Navarro-Gonz{\'{a}}lez, Font, Zhang, Lund, Oxborrow,
  Brandt, Caballero-Garc{\'{i}}a, Carrasco-Garc{\'{i}}a, Castell{\'{o}}n,
  {Castro Tirado}, Christiansen, Eyles, Fern{\'{a}}ndez-Garc{\'{i}}a, Genov,
  Guziy, Hu, {Nicuesa Guelbenzu}, Pandey, Peng, {P{\'{e}}rez del Pulgar},
  {Reina Terol}, Rodr{\'{i}}guez, S{\'{a}}nchez-Ram{\'{i}}rez, Sun, Ullaland,
  \& Yang}]{Castro-Tirado2021}
Castro-Tirado, A.~J., {\O}stgaard, N., G{\"{o}}ǧ{\"{u}}ş, E., {et~al.} 2021,
  Nature, 600, 621, \dodoi{10.1038/s41586-021-04101-1}

\bibitem[{Cenko {et~al.}(2010)Cenko, Butler, Ofek, Perley, Morgan, Frail,
  Gorosabel, Bloom, Castro-Tirado, Cepa, Chandra, {De Ugarte Postigo},
  Filippenko, Klein, Kulkarni, Miller, Nugent, \& Starr}]{Cenko2010}
Cenko, S.~B., Butler, N.~R., Ofek, E.~O., {et~al.} 2010, Astronomical Journal,
  140, 224, \dodoi{10.1088/0004-6256/140/1/224}

\bibitem[{Colaiuda {et~al.}(2009)Colaiuda, Beyer, \& Kokkotas}]{Colaiuda2009}
Colaiuda, A., Beyer, H., \& Kokkotas, K.~D. 2009, Monthly Notices of the Royal
  Astronomical Society, 396, 1441, \dodoi{10.1111/j.1365-2966.2009.14878.x}

\bibitem[{Colaiuda \& Kokkotas(2011)}]{Colaiuda2011}
Colaiuda, A., \& Kokkotas, K.~D. 2011, Monthly Notices of the Royal
  Astronomical Society, 414, 3014, \dodoi{10.1111/j.1365-2966.2011.18602.x}

\bibitem[{Cook {et~al.}(2006)Cook, Gelman, \& Rubin}]{Cook2006}
Cook, S.~R., Gelman, A., \& Rubin, D.~B. 2006, Journal of Computational and
  Graphical Statistics, 15, 675, \dodoi{10.1198/106186006X136976}

\bibitem[{Covino {et~al.}(2020)Covino, Landoni, Sandrinelli, \&
  Treves}]{Covino2020}
Covino, S., Landoni, M., Sandrinelli, A., \& Treves, A. 2020, The Astrophysical
  Journal, 895, 122, \dodoi{10.3847/1538-4357/ab8bd4}

\bibitem[{Czekala {et~al.}(2017)Czekala, Mandel, Andrews, Dittmann, Ghosh,
  Montet, \& Newton}]{Czekala2017}
Czekala, I., Mandel, K.~S., Andrews, S.~M., {et~al.} 2017, The Astrophysical
  Journal, 840, 49, \dodoi{10.3847/1538-4357/aa6aab}

\bibitem[{Davenport {et~al.}(2014)Davenport, Hawley, Hebb, Wisniewski,
  Kowalski, Johnson, Malatesta, Peraza, Keil, Silverberg, Jansen, Scheffler,
  Berdis, Larsen, \& Hilton}]{Davenport2014}
Davenport, J.~R., Hawley, S.~L., Hebb, L., {et~al.} 2014, Astrophysical
  Journal, 797, \dodoi{10.1088/0004-637X/797/2/122}

\bibitem[{{De Luca} {et~al.}(2010){De Luca}, Esposito, Israel, G{\"{o}}tz,
  Novara, Tiengo, \& Mereghetti}]{DeLuca2010}
{De Luca}, A., Esposito, P., Israel, G.~L., {et~al.} 2010, Monthly Notices of
  the Royal Astronomical Society, 402, 1870,
  \dodoi{10.1111/j.1365-2966.2009.16012.x}

\bibitem[{Delbridge {et~al.}(2019)Delbridge, Bindel, \& Wilson}]{Delbridge2019}
Delbridge, I.~A., Bindel, D.~S., \& Wilson, A.~G. 2019, arXiv.
\newblock \doarXiv{1912.12834}

\bibitem[{D'Emilio {et~al.}(2021)D'Emilio, Green, \& Raymond}]{DEmilio2021}
D'Emilio, V., Green, R., \& Raymond, V. 2021, Monthly Notices of the Royal
  Astronomical Society, 508, 2090, \dodoi{10.1093/mnras/stab2623}

\bibitem[{Dichiara {et~al.}(2013)Dichiara, Guidorzi, Frontera, \&
  Amati}]{Dichiara2013}
Dichiara, S., Guidorzi, C., Frontera, F., \& Amati, L. 2013, Astrophysical
  Journal, 777, \dodoi{10.1088/0004-637X/777/2/132}

\bibitem[{Dominique {et~al.}(2018)Dominique, Zhukov, Dolla, Inglis, \&
  Lapenta}]{Dominique2018}
Dominique, M., Zhukov, A.~N., Dolla, L., Inglis, A., \& Lapenta, G. 2018, Solar
  Physics, 293, 61, \dodoi{10.1007/s11207-018-1281-x}

\bibitem[{Eyer \& Bartholdi(1999)}]{Eyer1999}
Eyer, L., \& Bartholdi, P. 1999, Astronomy and Astrophysics Supplement Series,
  135, 1, \dodoi{10.1051/aas:1999102}

\bibitem[{Flaxman {et~al.}(2015)Flaxman, Wilson, Neill, Nickisch, \&
  Smola}]{Flaxman2015}
Flaxman, S., Wilson, A.~G., Neill, D.~B., Nickisch, H., \& Smola, A.~J. 2015,
  32nd International Conference on Machine Learning, ICML 2015, 1, 607

\bibitem[{Foreman-Mackey(2016)}]{Foreman-Mackey2016}
Foreman-Mackey, D. 2016, The Journal of Open Source Software, 1, 24,
  \dodoi{10.21105/joss.00024}

\bibitem[{Foreman-Mackey {et~al.}(2017)Foreman-Mackey, Agol, Ambikasaran, \&
  Angus}]{Foreman-Mackey2017}
Foreman-Mackey, D., Agol, E., Ambikasaran, S., \& Angus, R. 2017, The
  Astronomical Journal, 154, 220, \dodoi{10.3847/1538-3881/aa9332}

\bibitem[{Gabler {et~al.}(2011)Gabler, Cerd{\'{a}}-Dur{\'{a}}n, Font,
  M{\"{u}}ller, \& Stergioulas}]{Gabler2011}
Gabler, M., Cerd{\'{a}}-Dur{\'{a}}n, P., Font, J.~A., M{\"{u}}ller, E., \&
  Stergioulas, N. 2011, Monthly Notices of the Royal Astronomical Society:
  Letters, 410, L37, \dodoi{10.1111/j.1745-3933.2010.00974.x}

\bibitem[{Gabler {et~al.}(2013)Gabler, Cerda-Duran, Font, Muller, \&
  Stergioulas}]{Gabler2013}
Gabler, M., Cerda-Duran, P., Font, J.~A., Muller, E., \& Stergioulas, N. 2013,
  Monthly Notices of the Royal Astronomical Society, 430, 1811,
  \dodoi{10.1093/mnras/sts721}

\bibitem[{Gardner {et~al.}(2018)Gardner, Pleiss, Bindel, Weinberger, \&
  Wilson}]{Gardner2018}
Gardner, J.~R., Pleiss, G., Bindel, D., Weinberger, K.~Q., \& Wilson, A.~G.
  2018, Advances in Neural Information Processing Systems, 2018-Decem, 7576.
\newblock \doarXiv{1809.11165}

\bibitem[{Gentle(2009)}]{Gentle2009}
Gentle, J.~E. 2009, {Computational Statistics}, Statistics and Computing (New
  York, NY: Springer New York), \dodoi{10.1007/978-0-387-98144-4}

\bibitem[{Gibson {et~al.}(2012)Gibson, Aigrain, Roberts, Evans, Osborne, \&
  Pont}]{Gibson2012}
Gibson, N.~P., Aigrain, S., Roberts, S., {et~al.} 2012, Monthly Notices of the
  Royal Astronomical Society, 419, 2683,
  \dodoi{10.1111/j.1365-2966.2011.19915.x}

\bibitem[{Grechnev {et~al.}(2003)Grechnev, White, \& Kundu}]{Grechnev2003}
Grechnev, V.~V., White, S.~M., \& Kundu, M.~R. 2003, The Astrophysical Journal,
  588, 1163, \dodoi{10.1086/374315}

\bibitem[{Grunblatt {et~al.}(2015)Grunblatt, Howard, \&
  Haywood}]{Grunblatt2015}
Grunblatt, S.~K., Howard, A.~W., \& Haywood, R.~D. 2015, The Astrophysical
  Journal, 808, 127, \dodoi{10.1088/0004-637X/808/2/127}

\bibitem[{Guidorzi {et~al.}(2016)Guidorzi, Dichiara, \& Amati}]{Guidorzi2016}
Guidorzi, C., Dichiara, S., \& Amati, L. 2016, Astronomy and Astrophysics, 589,
  1, \dodoi{10.1051/0004-6361/201527642}

\bibitem[{Hayes {et~al.}(2019)Hayes, Gallagher, Dennis, Ireland, Inglis, \&
  Morosan}]{Hayes2019}
Hayes, L.~A., Gallagher, P.~T., Dennis, B.~R., {et~al.} 2019, arXiv,
  \dodoi{10.3847/1538-4357/ab0ca3}

\bibitem[{Hayes {et~al.}(2016)Hayes, Gallagher, Dennis, Ireland, Inglis, \&
  Ryan}]{Hayes2016}
---. 2016, The Astrophysical Journal, 827, L30,
  \dodoi{10.3847/2041-8205/827/2/l30}

\bibitem[{Hayes {et~al.}(2020)Hayes, Inglis, Christe, Dennis, \&
  Gallagher}]{Hayes2020}
Hayes, L.~A., Inglis, A.~R., Christe, S., Dennis, B., \& Gallagher, P.~T. 2020,
  The Astrophysical Journal, 895, 50, \dodoi{10.3847/1538-4357/ab8d40}

\bibitem[{Huebner {et~al.}(2021)Huebner, Huppenkothen, Lasky, \&
  Inglis}]{Huebner2021}
Huebner, M., Huppenkothen, D., Lasky, P.~D., \& Inglis, A.~R. 2021.
\newblock \doarXiv{2108.07418}

\bibitem[{Huppenkothen \& Bachetti(2021)}]{Huppenkothen2021}
Huppenkothen, D., \& Bachetti, M. 2021, 19, 1.
\newblock \doarXiv{2104.03278}

\bibitem[{Huppenkothen {et~al.}(2014{\natexlab{a}})Huppenkothen, Heil, Watts,
  \& G{\"{o}}ğ{\"{u}}ş}]{Huppenkothen2014b}
Huppenkothen, D., Heil, L.~M., Watts, A.~L., \& G{\"{o}}ğ{\"{u}}ş, E.
  2014{\natexlab{a}}, Astrophysical Journal, 795,
  \dodoi{10.1088/0004-637X/795/2/114}

\bibitem[{Huppenkothen {et~al.}(2014{\natexlab{b}})Huppenkothen, Watts, \&
  Levin}]{Huppenkothen2014a}
Huppenkothen, D., Watts, A.~L., \& Levin, Y. 2014{\natexlab{b}}, Astrophysical
  Journal, 793, \dodoi{10.1088/0004-637X/793/2/129}

\bibitem[{Huppenkothen {et~al.}(2012)Huppenkothen, Watts, Uttley, van~der
  Horst, van~der Klis, Kouveliotou, Gogus, Granot, Vaughan, \&
  Finger}]{Huppenkothen2012}
Huppenkothen, D., Watts, A.~L., Uttley, P., {et~al.} 2012, Astrophysical
  Journal, 768, \dodoi{10.1088/0004-637X/768/1/87}

\bibitem[{Huppenkothen {et~al.}(2014{\natexlab{c}})Huppenkothen, D'Angelo,
  Watts, Heil, van~der Klis, van~der Horst, Kouveliotou, Baring, Gogus, Granot,
  Kaneko, Lin, von Kienlin, \& Younes}]{Huppenkothen2014}
Huppenkothen, D., D'Angelo, C., Watts, A.~L., {et~al.} 2014{\natexlab{c}},
  Astrophysical Journal, 787, \dodoi{10.1088/0004-637X/787/2/128}

\bibitem[{Huppenkothen {et~al.}(2015)Huppenkothen, Brewer, Hogg, Murray, Frean,
  Elenbaas, Watts, Levin, van~der Horst, \& Kouveliotou}]{Huppenkothen2015}
Huppenkothen, D., Brewer, B.~J., Hogg, D.~W., {et~al.} 2015, Astrophysical
  Journal, 810, \dodoi{10.1088/0004-637X/810/1/66}

\bibitem[{Huppenkothen {et~al.}(2017)Huppenkothen, Younes, Ingram, Kouveliotou,
  G{\"{o}}ğ{\"{u}}ş, Bachetti, S{\'{a}}nchez-Fern{\'{a}}ndez, Chenevez,
  Motta, van~der Klis, Granot, Gehrels, Kuulkers, Tomsick, \&
  Walton}]{Huppenkothen2017}
Huppenkothen, D., Younes, G., Ingram, A., {et~al.} 2017, The Astrophysical
  Journal, 834, 90, \dodoi{10.3847/1538-4357/834/1/90}

\bibitem[{Hurley {et~al.}(2005)Hurley, Boggs, Smith, Duncan, Lin, Zoglauer,
  Krucker, Hurford, Hudson, Wigger, Hajdas, Thompson, Mitrofanov, Sanin,
  Boynton, Fellows, von Kienlin, Lichti, Rau, \& Cline}]{Hurley2005}
Hurley, K., Boggs, S.~E., Smith, D.~M., {et~al.} 2005, Nature, 434, 1098,
  \dodoi{10.1038/nature03519}

\bibitem[{Inglis {et~al.}(2016)Inglis, Ireland, Dennis, Hayes, \&
  Gallagher}]{Inglis2016}
Inglis, A.~R., Ireland, J., Dennis, B.~R., Hayes, L., \& Gallagher, P. 2016,
  The Astrophysical Journal, 833, 284, \dodoi{10.3847/1538-4357/833/2/284}

\bibitem[{Inglis {et~al.}(2015)Inglis, Ireland, \& Dominique}]{Inglis2015}
Inglis, A.~R., Ireland, J., \& Dominique, M. 2015, The Astrophysical Journal,
  798, 108, \dodoi{10.1088/0004-637X/798/2/108}

\bibitem[{Inglis {et~al.}(2008)Inglis, Nakariakov, \& Melnikov}]{Inglis2008}
Inglis, A.~R., Nakariakov, V.~M., \& Melnikov, V.~F. 2008, Astronomy and
  Astrophysics, 487, 1147, \dodoi{10.1051/0004-6361:20079323}

\bibitem[{Ingram \& Motta(2019)}]{Ingram2019}
Ingram, A.~R., \& Motta, S.~E. 2019, New Astronomy Reviews, 85, 101524,
  \dodoi{10.1016/j.newar.2020.101524}

\bibitem[{Israel {et~al.}(2005)Israel, Belloni, Stella, Rephaeli, Gruber,
  Casella, Dall'Osso, Rea, Persic, \& Rothschild}]{Israel2005}
Israel, G.~L., Belloni, T., Stella, L., {et~al.} 2005, The Astrophysical
  Journal, 628, L53, \dodoi{10.1086/432615}

\bibitem[{Iwakiri {et~al.}(2010)Iwakiri, Ohno, Kamae, Nakagawa, Terada,
  Tashiro, Yoshida, Yamaoka, \& Makishima}]{Iwakiri2010}
Iwakiri, W., Ohno, M., Kamae, T., {et~al.} 2010, AIP Conference Proceedings,
  1279, 89, \dodoi{10.1063/1.3509358}

\bibitem[{Kane {et~al.}(1983)Kane, Kai, Kosugi, Enome, Landecker, \&
  McKenzie}]{Kane1983}
Kane, S.~R., Kai, K., Kosugi, T., {et~al.} 1983, The Astrophysical Journal,
  271, 376, \dodoi{10.1086/161203}

\bibitem[{Kelly {et~al.}(2014)Kelly, Becker, Sobolewska, Siemiginowska, \&
  Uttley}]{Kelly2014}
Kelly, B.~C., Becker, A.~C., Sobolewska, M., Siemiginowska, A., \& Uttley, P.
  2014, The Astrophysical Journal, 788, 33, \dodoi{10.1088/0004-637X/788/1/33}

\bibitem[{Kupriyanova {et~al.}(2016)Kupriyanova, Kashapova, Reid, \&
  Myagkova}]{Kupriyanova2016}
Kupriyanova, E.~G., Kashapova, L.~K., Reid, H. A.~S., \& Myagkova, I.~N. 2016,
  Solar Physics, 291, 3427, \dodoi{10.1007/s11207-016-0958-2}

\bibitem[{Levin(2006)}]{Levin2006}
Levin, Y. 2006, Monthly Notices of the Royal Astronomical Society: Letters,
  368, \dodoi{10.1111/j.1745-3933.2006.00155.x}

\bibitem[{Levin(2007)}]{Levin2007}
---. 2007, Monthly Notices of the Royal Astronomical Society, 377, 159,
  \dodoi{10.1111/j.1365-2966.2007.11582.x}

\bibitem[{Levin \& van Hoven(2011)}]{Levin2011}
Levin, Y., \& van Hoven, M. 2011, Monthly Notices of the Royal Astronomical
  Society, 418, 659, \dodoi{10.1111/j.1365-2966.2011.19515.x}

\bibitem[{{Lindberg} {et~al.}(2022){Lindberg}, {Huppenkothen}, {Jones},
  {Bolin}, {Juri{\'c}}, {Golkhou}, {Bellm}, {Drake}, {Graham}, {Laher},
  {Mahabal}, {Masci}, {Riddle}, \& {Shin}}]{willeckelindberg2022}
{Lindberg}, C.~W., {Huppenkothen}, D., {Jones}, R.~L., {et~al.} 2022, \aj, 163,
  29, \dodoi{10.3847/1538-3881/ac3079}

\bibitem[{{Luger} {et~al.}(2021){Luger}, {Foreman-Mackey}, \&
  {Hedges}}]{luger2021}
{Luger}, R., {Foreman-Mackey}, D., \& {Hedges}, C. 2021, \aj, 162, 124,
  \dodoi{10.3847/1538-3881/abfdb9}

\bibitem[{Masada {et~al.}(2007)Masada, Kawanaka, Sano, \& Shibata}]{Masada2007}
Masada, Y., Kawanaka, N., Sano, T., \& Shibata, K. 2007, The Astrophysical
  Journal, 663, 437, \dodoi{10.1086/518088}

\bibitem[{Massey(1951)}]{Massey1951}
Massey, F.~J. 1951, Journal of the American Statistical Association, 46, 68,
  \dodoi{10.2307/2280095}

\bibitem[{McLaughlin {et~al.}(2018)McLaughlin, Nakariakov, Dominique,
  Jel{\'{i}}nek, \& Takasao}]{McLaughlin2018}
McLaughlin, J.~A., Nakariakov, V.~M., Dominique, M., Jel{\'{i}}nek, P., \&
  Takasao, S. 2018, Space Science Reviews, 214, 1,
  \dodoi{10.1007/s11214-018-0478-5}

\bibitem[{Melnikov {et~al.}(2005)Melnikov, Reznikova, Shibasaki, \&
  Nakariakov}]{Melnikov2005}
Melnikov, V.~F., Reznikova, V.~E., Shibasaki, K., \& Nakariakov, V.~M. 2005,
  Astronomy \& Astrophysics, 439, 727, \dodoi{10.1051/0004-6361:20052774}

\bibitem[{Miller {et~al.}(2019)Miller, Chirenti, \& Strohmayer}]{Miller2019}
Miller, M.~C., Chirenti, C., \& Strohmayer, T.~E. 2019, The Astrophysical
  Journal, 871, 95, \dodoi{10.3847/1538-4357/aaf5ce}

\bibitem[{Moore {et~al.}(2016)Moore, Berry, Chua, \& Gair}]{Moore2016}
Moore, C.~J., Berry, C. P.~L., Chua, A. J.~K., \& Gair, J.~R. 2016, Physical
  Review D, 93, 064001, \dodoi{10.1103/PhysRevD.93.064001}

\bibitem[{Nakariakov {et~al.}(2010)Nakariakov, Foullon, Myagkova, \&
  Inglis}]{Nakariakov2010}
Nakariakov, V.~M., Foullon, C., Myagkova, I.~N., \& Inglis, A.~R. 2010, The
  Astrophysical Journal, 708, L47, \dodoi{10.1088/2041-8205/708/1/L47}

\bibitem[{Nakariakov \& Melnikov(2009)}]{Nakariakov2009}
Nakariakov, V.~M., \& Melnikov, V.~F. 2009, Space Science Reviews, 149, 119,
  \dodoi{10.1007/s11214-009-9536-3}

\bibitem[{Norris {et~al.}(1996)Norris, Nemiroff, Bonnell, Scargle, Kouveliotou,
  Paciesas, Meegan, \& Fishman}]{Norris1996}
Norris, J.~P., Nemiroff, R.~J., Bonnell, J.~T., {et~al.} 1996, The
  Astrophysical Journal, 459, 393, \dodoi{10.1086/176902}

\bibitem[{Palmer {et~al.}(2005)Palmer, Barthelmy, Gehrels, Kippen, Cayton,
  Kouveliotou, Eichler, Wijers, Woods, Granot, Lyubarsky, Ramirez-Ruiz,
  Barbier, Chester, Cummings, Fenimore, Finger, Gaensler, Hullinger, Krimm,
  Markwardt, Nousek, Parsons, Patel, Sakamoto, Sato, Suzuki, \&
  Tueller}]{Palmer2005}
Palmer, D.~M., Barthelmy, S., Gehrels, N., {et~al.} 2005, Nature, 434, 1107,
  \dodoi{10.1038/nature03525}

\bibitem[{Parks \& Winckler(1969)}]{Parks1969}
Parks, G.~K., \& Winckler, J.~R. 1969, The Astrophysical Journal, 155, L117,
  \dodoi{10.1086/180315}

\bibitem[{Pastor-Marazuela {et~al.}(2022)Pastor-Marazuela, van Leeuwen, Bilous,
  Connor, Maan, Oostrum, Petroff, Straal, Vohl, Adams, Adebahr, Attema,
  Boersma, van~den Brink, van Cappellen, Coolen, Damstra, D{\'{e}}nes, Hess,
  van~der Hulst, Hut, Kutkin, Loose, Lucero, Mika, Moss, Mulder, Norden,
  Oosterloo, Rajwade, van~der Schuur, Sclocco, Smits, \&
  Ziemke}]{PastorMarazuela2022}
Pastor-Marazuela, I., van Leeuwen, J., Bilous, A., {et~al.} 2022, 1.
\newblock \doarXiv{2202.08002}

\bibitem[{Rasmussen \& Williams(2006)}]{Rasmussen2006}
Rasmussen, C., \& Williams, C. 2006, {Gaussian Processes for Machine Learning},
  Vol.~7 (MIT Press)

\bibitem[{Rea {et~al.}(2009)Rea, Israel, Turolla, Esposito, Mereghetti,
  G{\"{o}}tz, Zane, Tiengo, Hurley, Feroci, Still, Yershov, Winkler, Perna,
  Bernardini, Ubertini, Stella, Campana, van~der Klis, \& Woods}]{Rea2009}
Rea, N., Israel, G.~L., Turolla, R., {et~al.} 2009, Monthly Notices of the
  Royal Astronomical Society, 396, 2419,
  \dodoi{10.1111/j.1365-2966.2009.14920.x}

\bibitem[{Romero-Shaw {et~al.}(2020)Romero-Shaw, Talbot, Biscoveanu, D'Emilio,
  Ashton, Berry, Coughlin, Galaudage, Hoy, H{\"{u}}bner, Phukon, Pitkin, Rizzo,
  Sarin, Smith, Stevenson, Vajpeyi, Ar{\`{e}}ne, Athar, Banagiri, Bose, Carney,
  Chatziioannou, Clark, Colleoni, Cotesta, Edelman, Estell{\'{e}}s,
  Garc{\'{i}}a-Quir{\'{o}}s, Ghosh, Green, Haster, Husa, Keitel, Kim,
  Hernandez-Vivanco, {Maga{\~{n}}a Hernandez}, Karathanasis, Lasky, {De Lillo},
  Lower, Macleod, Mateu-Lucena, Miller, Millhouse, Morisaki, Oh, Ossokine,
  Payne, Powell, Pratten, P{\"{u}}rrer, Ramos-Buades, Raymond, Thrane, Veitch,
  Williams, Williams, \& Xiao}]{Romero-Shaw2020}
Romero-Shaw, I.~M., Talbot, C., Biscoveanu, S., {et~al.} 2020, Monthly Notices
  of the Royal Astronomical Society, 499, 3295, \dodoi{10.1093/mnras/staa2850}

\bibitem[{Samuelsson \& Andersson(2007)}]{Samuelsson2007}
Samuelsson, L., \& Andersson, N. 2007, Monthly Notices of the Royal
  Astronomical Society, 374, 256, \dodoi{10.1111/j.1365-2966.2006.11147.x}

\bibitem[{Savitzky \& Golay(1964)}]{Savitzky1964}
Savitzky, A., \& Golay, M. J.~E. 1964, Analytical Chemistry, 36, 1627,
  \dodoi{10.1021/ac60214a047}

\bibitem[{Skilling(2004)}]{Skilling2004}
Skilling, J. 2004, in AIP Conference Proceedings, Vol. 735 (AIP), 395--405,
  \dodoi{10.1063/1.1835238}

\bibitem[{Skilling(2006)}]{Skilling2006}
Skilling, J. 2006, Bayesian Analysis, 1, 833, \dodoi{10.1214/06-BA127}

\bibitem[{Smith {et~al.}(2020)Smith, Ashton, Vajpeyi, \& Talbot}]{Smith2020}
Smith, R.~J., Ashton, G., Vajpeyi, A., \& Talbot, C. 2020, Monthly Notices of
  the Royal Astronomical Society, 498, 4492, \dodoi{10.1093/mnras/staa2483}

\bibitem[{Sotani {et~al.}(2016)Sotani, Iida, \& Oyamatsu}]{Sotani2016}
Sotani, H., Iida, K., \& Oyamatsu, K. 2016, New Astronomy, 43, 80,
  \dodoi{10.1016/j.newast.2015.08.003}

\bibitem[{Sotani {et~al.}(2017)Sotani, Iida, \& Oyamatsu}]{Sotani2017}
---. 2017, Monthly Notices of the Royal Astronomical Society, 464, 3101,
  \dodoi{10.1093/mnras/stw2575}

\bibitem[{Sotani {et~al.}(2007)Sotani, Kokkotas, \& Stergioulas}]{Sotani2007}
Sotani, H., Kokkotas, K.~D., \& Stergioulas, N. 2007, Monthly Notices of the
  Royal Astronomical Society, 375, 261,
  \dodoi{10.1111/j.1365-2966.2006.11304.x}

\bibitem[{Sotani {et~al.}(2008)Sotani, Kokkotas, \& Stergioulas}]{Sotani2008}
---. 2008, Monthly Notices of the Royal Astronomical Society: Letters, 385, L5,
  \dodoi{10.1111/j.1745-3933.2007.00420.x}

\bibitem[{{Speagle}(2020)}]{speagle2020}
{Speagle}, J.~S. 2020, \mnras, 493, 3132, \dodoi{10.1093/mnras/staa278}

\bibitem[{Strohmayer \& Watts(2005)}]{Strohmayer2005}
Strohmayer, T.~E., \& Watts, A.~L. 2005, The Astrophysical Journal, 632, L111,
  \dodoi{10.1086/497911}

\bibitem[{Strohmayer \& Watts(2006)}]{Strohmayer2006}
---. 2006, The Astrophysical Journal, 653, 593, \dodoi{10.1086/508703}

\bibitem[{Tarnopolski \& Marchenko(2021)}]{Tarnopolski2021}
Tarnopolski, M., \& Marchenko, V. 2021, The Astrophysical Journal, 911, 20,
  \dodoi{10.3847/1538-4357/abe5b1}

\bibitem[{Timmer \& K{\"{o}}nig(1995)}]{Timmer1995}
Timmer, J., \& K{\"{o}}nig, M. 1995, Astronomy and Astrophysics, 300, 707

\bibitem[{{Van Doorsselaere} {et~al.}(2016){Van Doorsselaere}, Kupriyanova, \&
  Yuan}]{VanDoorsselaere2016}
{Van Doorsselaere}, T., Kupriyanova, E.~G., \& Yuan, D. 2016, Solar Physics,
  291, 3143, \dodoi{10.1007/s11207-016-0977-z}

\bibitem[{van Haasteren \& Vallisneri(2014)}]{VanHaasteren2014}
van Haasteren, R., \& Vallisneri, M. 2014, Physical Review D, 90, 104012,
  \dodoi{10.1103/PhysRevD.90.104012}

\bibitem[{VanderPlas(2018)}]{VanderPlas2018}
VanderPlas, J.~T. 2018, The Astrophysical Journal Supplement Series, 236, 16,
  \dodoi{10.3847/1538-4365/aab766}

\bibitem[{Wandelt \& Hansen(2003)}]{Wandelt2003}
Wandelt, B.~D., \& Hansen, F.~K. 2003, Physical Review D, 67, 023001,
  \dodoi{10.1103/PhysRevD.67.023001}

\bibitem[{Watts(2012)}]{Watts2012a}
Watts, A.~L. 2012, Annual Review of Astronomy and Astrophysics, 50, 609,
  \dodoi{10.1146/annurev-astro-040312-132617}

\bibitem[{Watts \& Strohmayer(2006)}]{Watts2006}
Watts, A.~L., \& Strohmayer, T.~E. 2006, The Astrophysical Journal, 637, L117,
  \dodoi{10.1086/500735}

\bibitem[{Wilson \& Nickisch(2015)}]{Wilson2015}
Wilson, A.~G., \& Nickisch, H. 2015, 32nd International Conference on Machine
  Learning, ICML 2015, 3, 1775.
\newblock \doarXiv{1503.01057}

\bibitem[{Zhu \& Thrane(2020)}]{Zhu2020}
Zhu, X.~J., \& Thrane, E. 2020, arXiv, \dodoi{10.3847/1538-4357/abac5a}

\bibitem[{Ziaeepour \& Gardner(2011)}]{Ziaeepour2011}
Ziaeepour, H., \& Gardner, B. 2011, Journal of Cosmology and Astroparticle
  Physics, 2011, \dodoi{10.1088/1475-7516/2011/12/001}

\bibitem[{Zimovets {et~al.}(2021)Zimovets, McLaughlin, Srivastava, Kolotkov,
  Kuznetsov, Kupriyanova, Cho, Inglis, Reale, Pascoe, Tian, Yuan, Li, \&
  Zhang}]{Zimovets2021}
Zimovets, I.~V., McLaughlin, J.~A., Srivastava, A.~K., {et~al.} 2021, Space
  Science Reviews, 217, 66, \dodoi{10.1007/s11214-021-00840-9}

\end{thebibliography}
\end{document}